\documentclass{JHEP3}
\usepackage{amsmath,amsfonts,amssymb,amsthm,amstext,eucal}
\usepackage{graphicx,lscape}
\def\bea{\begin{eqnarray}}
\def\eea{\end{eqnarray}}
\def\beann{\begin{eqnarray*}}
\def\eeann{\end{eqnarray*}}
\def\be{\begin{equation}}
\def\ee{\end{equation}}
\def\ba{\begin{array}}
\def\ea{\end{array}}
\def\ben{\begin{enumerate}}
\def\een{\end{enumerate}}
\def\4{\tilde }
\def\5{\bar }
\def\6{\partial }
\def\7{\hat }

\def\bz{\bar{z}}
\def\bnu{\bar{\nu}}
\def\bmu{\bar{\mu}}

\def\cL{{\cal L}}
\def\cM{{\cal M}}
\def\cN{{\cal N}}

\def\cR{{\cal R}}
\font\mybb=msbm10 at 10pt
\def\bb#1{\hbox{\mybb#1}}
\def\bR {\bb{R}}
\def\bZ {\bb{Z}}
\def\bE {\bb{E}}

\def\bC {\bb{C}}
\def\sp{\partial\!\!\!\!/}

\newcommand{\fsu}{\mathfrak{su}}

\preprint{hep-th/0310255 \\
ITFA-2003-51 \\
UPR-1054-T\\
WIS/28/03-OCT-DPP
}
\title{Killing spectroscopy of closed timelike curves}
\author{
Liat Maoz$^{1}$ and Joan Sim\'{o}n$^{2,3}$\\
~\\
1. Institute for Theoretical Physics, University of Amsterdam\\
 Valckenierstraat 65, 1018XE Amsterdam, The Netherlands \\
~\\
2. Department of Physics and Astronomy, David Rittenhouse Laboratory\\
The University of Pennsylvania \\
209 South 33rd Street, Philadelphia, PA 19104-6396, U.S.A.\\
~\\
3. Department of Particle Physics, The Weizmann Institute of Science\\
Herzl Street 2, 76100 Rehovot, Israel\\
~\\

\email{lmaoz@science.uva.nl, jsimon@bokchoy.hep.upenn.edu} }

\abstract{We analyse the existence of closed timelike curves in spacetimes which possess an isometry. In
particular we check which discrete quotients of such spaces lead to closed timelike curves. As
a by-product of our analysis, we prove that the notion of existence or non-existence of closed timelike curves is
a T-duality invariant notion, whenever the direction along which we apply such transformations is everywhere spacelike.
Our formalism is straightforwardly applied to supersymmetric theories. We provide some new examples in the
context of D-branes and generalized pp-waves.}

\keywords{Closed timelike curves, (super)symmetry, dualities, discrete quotients}

\begin{document}

\section{Introduction and summary}
\label{sec:intro}

It is by now well-known that there exist solutions of general relativity which are locally causal, but have closed
timelike curves (CTCs) on a global scale \cite{visser}. This is the case both in the presence and absence of
curvature singularities. Needless to say the existence of such curves is considered to be problematic and the traditional
attitude has been to consider the subset of spacetimes containing them as non-physical.

Lately, spacetimes which violate the chronology condition received renewed attention by the string theory community after
it was realized in \cite{firstgodel} that G\"{o}del-like universes \cite{godel} can be supersymmetrically embedded in string
theory. It was later realized and emphasized that these solutions were T-dual to compactified plane wave backgrounds
\cite{Horava,Harmark}. In \cite{Horava}, the issue of the non-physical character of these spacetimes was re-examined.
It was suggested that holography might provide new insights on this subject. A similar philosophy was advocated in \cite{Rozali}.
On the other hand, in \cite{Taming}, there was an attempt to define field theory in a non-globally hyperbolic spacetime having
CTCs based on standard orbifold ideas in string theory.

As string theorists, we do not have currently any fundamental principle to select or discard a priori any vacuum.
Thus, it is our belief that one should understand the dynamics in these classical spacetimes violating the
chronology condition before deciding their fate. Following this direction, it was shown in \cite{prl} that by probing
one of these spacetimes with an adequate probe, a supertube \cite{supertubes}, the metric on moduli space
develops a singularity. A similar phenomenum was reported in the enhan\c{c}on mechanism \cite{enhancon}. These
suggested to replace the metric in the chronology violating region by one describing a domain wall of supertubes
which is free of CTCs. Recently, it was reported in \cite{dan} that a similar phenomenum is happening for
G\"{o}del/$AdS_3$, where the probes are now long strings. The supertube domain wall seems to provide a more
natural framework to discuss holography, and this was the direction followed in \cite{ustwo}, where the connection
between the physics in G\"{o}del-type universes and the Landau problem pointed out in \cite{rey} played an
important role\footnote{There have been other works in the recent literature concerning the physics of
G\"{o}del-type universes in their different dual incarnations. See \cite{others,carlos1} and references therein.}.

In order to investigate how universal these mechanisms might be, or simply to increase our knowledge on the subject,
it should be of interest to determine whether CTCs exist or are absent in arbitrary backgrounds. In this note, we will make
some more precise statements regarding this topic. To do so, we shall deal with spacetimes which are time-orientable,
and which possess some isometry with a definite sign norm, i.e. we shall deal with spacetimes having a Killing vector which is everywhere
timelike, everywhere lightlike or everywhere spacelike. First, we shall look at such spacetimes where the Killing
direction is non-compact and try to identify the conditions for CTCs to exist. We shall afterwards look at the
possible discrete quotients of these spaces and try to characterize which of these have CTCs. Incidentally, we would like
to stress that our organizing symmetry principle (a standard technique in general relativity) is also singled out by supersymmetry
considerations. Indeed, assuming the existence of a single supersymmetry, it is known \cite{jerome} that
there should exist causal isometries in the metric. If there is more than one supersymmetry, one can also have
spacelike isometries. These simple ideas have been developed to attempt a classification of all supersymmetric supergravity
backgrounds in different dimensions \cite{firstgodel, jerome, reall}.

A second important goal of this note is to analyse whether the existence or non-existence of CTCs is a notion invariant under
dualities. One may think that in certain scenarios, either strong coupling effects or specific stringy features such as
the existence of T-duality, may help in resolving causal puzzles. T-duality might be particularly confusing due to the fact
that ``metric information'' is being exchanged with ``flux information'', and so it is a priori unclear how this would affect the existence of CTCs.
We shall prove that none of these possibilities are realized in the supergravity regime description. The existence or non-existence
of CTCs is a \emph{U-duality invariant notion}, i.e. there are no U-duality transformations taking us from a spacetime which
has CTCs into one not having them, under certain assumptions to be discussed in the body of the paper. Therefore, if there is
any kind of instability in the dynamics of classical supergravity configurations having CTCs, it should also be manifest
in their U-dual descriptions.

In spite of the fact that symmetric backgrounds are not the most general ones we could have considered, (even
though some of our techniques do apply to more generic backgrounds), it should be clear that our analysis is
relevant in many physically interesting set-ups. The existence of spacelike isometries, and in particular, compact
and periodic ones is intrinsically tied to Kaluza--Klein reductions and compactifications in general. PP-wave
backgrounds are examples of spacetimes having null Killing vectors. Actually pp-wave backgrounds have a
covariantly null Killing vector, and in this sense are just special cases of the more general backgrounds with a
null Killing vector considered in this note. It has been shown recently that for pp-waves, with flat transverse
space, and a non-compact $u$-direction there are no CTCs, yet certain compactifications of these spaces do create
CTCs \cite{Sanchez, Carlos2,Mukund} and are T-dual to G\"{o}del type universes \cite{Horava,Harmark}. Finally, all
stationary configurations have a timelike Killing vector. Among these are both static configurations, and more
interestingly for our purposes, rotating ones \cite{george99}. In some of these cases, there is a rotational
parameter which above a certain threshold value gives 'over-rotating' geometries which contain CTCs. For example,
the BMPV rotating black hole which for large enough angular momentum $J$ develops CTCs outside the horizon
\cite{BMPV,aboutBMPV}. In particular, it was pointed out in \cite{carlos1,aboutBMPV} that the dual CFT corresponding to
these over-rotating geometries would be non-unitary. Similar black hole solutions with G\"{o}del universe
asymptotics were recently discussed in \cite{Carlos2,BHGoedl}. Another example is the supertubes which can be thought of
as a Myers blow up of rotating D0s and F1s into a D2. For large values of the angular momentum, one obtains
spacetimes with CTCs \cite{supertubes1}. A similar phenomenum happens for the M-theory partners of the supertubes
\cite{ribbons}. It is only when this system is built from a microscopic picture, dual to
an oscillating string carrying travelling waves, that such CTCs never appear \cite{ourHelF1}. Dual pictures as the
rotating D1-D5 system and other systems as the null scissors \cite{Bachas} also fall into the same category.

Let us now briefly summarize the main results of our paper. For spacetimes possessing a timelike isometry $\partial_t$, we
have shown that for non-compact $t$, the spacetime can have CTCs and have written down the conditions for their existence.
G\"{o}del-like universes belong to this class. Discrete quotients of the latter by $\partial_t$ or by $\partial_t+\beta\partial_\psi$
where $\psi$ is a periodic spacelike isometry of the background always create CTCs. These results are shown in section \ref{sec:time}.
For spacetimes with a null isometry $\partial_v$, and a non-compact $u$ coordinate, there cannot be CTCs. The same statement applies
to discrete quotients generated by $\partial_v + \gamma\,V$, $V$ standing for an isometry of the configuration. On the other hand,
discrete quotients generated by $\partial_u + \beta\,\partial_v + \gamma\,V$ can give rise to spacetimes having CTCs under certain circumstances
that are spelled out in section \ref{sec:null}. The compactified plane-waves which are T-dual to G\"{o}del-like universes belong to this class.
For spacetimes with a spacelike isometry $\partial_z$, we have shown that a discrete quotient along
$\partial_z$ creates CTCs iff the 9-dimensional space obtained from the Kaluza--Klein reduction along $\partial_z$ also has CTCs.
The spacelike case is analysed in section \ref{sec:space}. As a by-product of this analysis, we prove in section
\ref{sec:ctcdual} that the existence or non-existence of CTCs is a T-duality invariant notion whenever the direction
along we consider this transformation is everywhere spacelike. We comment on the T-duality relations between the different
spacetimes discussed before, generalising the known compactified pp-wave relation to G\"{o}del-like universes to
adequate discrete quotients of pp-waves propagating in curved spacetimes.

In section 3, we work out a few interesting examples of supersymmetric IIB backgrounds which have a timelike or null isometry
and no CTCs. We show how certain discrete quotients by an everywhere spacelike
Killing vector can create spacetimes which have CTCs. In
particular we work out an example of a D3 brane background quotiented by a timelike isometry combined with a rotation transverse
 to the branes, and show CTCs are created by the quotient.
We work out the case of the supersymmetric IIB pp-wave backgrounds described in \cite{juanpp} with flat transverse
 space and show how a quotient by the null isometry combined with $\partial_u$ and a rotation creates CTCs. We also show their T-duals
give G\"{o}del type universes. Then we work out the case when the transverse space is an Eguchi-Hanson and perform a similar
quotient. We obtain metrics with CTCs, and write down their T-duals which would be the Eguchi-Hanson version of the
G\"{o}del type universes (GEH universes). Finally we work out the case of supersymmetric backgrounds in type IIB that consist of
pp-wave backgrounds considered in \cite{juanpp} superimposed in a supersymmetric way with D-branes. We show that quotienting as
before we get interesting spacetimes with CTCs.

Some of the more technical aspects of our work are described as appendices appearing at the end of this note. In particular, in
appendix \ref{sec:appA} we prove that the future directness of a given timelike vector field is independent of the choice of the
representative of the class of mutually future directed globally defined timelike vector fields. In appendix \ref{sec:appB}, we prove that
given a ten dimensional spacetime with a compact and periodic spacelike isometry having a future directed CTC, then its nine dimensional
Kaluza--Klein reduced spacetime has, \emph{necessarily}, a future directed CTC. In appendix \ref{sec:appC}, we analyse which Killing vectors
of the pp-wave backgrounds discussed in \cite{juanpp} preserve their supersymmetries when viewed as generators
of a discrete quotient. Finally, in appendix \ref{sec:appD}, we construct new supersymmetric type IIB supergravity configurations consisting of the pp-waves
in \cite{juanpp} superimposed with D-strings. Using U-duality and liftings to M-theory, other brane and pp-wave backgrounds of types IIA, IIB or
M-theory can be trivially obtained.

\section{On closed timelike curves in symmetric backgrounds}
\label{sec:ctcsymm}

We would like to study the kind of spacetimes that have closed timelike curves. Since their existence or non-existence
is only a metric dependent question, we shall focus only on this piece of information coming from the classical
supergravity configuration. When discussing whether such a feature is U-duality invariant, we shall include the fluxes.
We shall use as our organizing principle the existence of global symmetries in the spacetimes under consideration. We will therefore
focus on cases where the geometry possesses some Killing vector $k$, and in particular, on those in which $k$ is everywhere timelike,
everywhere null or everywhere spacelike.

Assuming the existence of such symmetries, one can write a local description for their metric geometries in one of the
following forms:
\begin{itemize}
  \item[(i)] $g(k\,,k)<0$, corresponding to a \emph{timelike} Killing vector field
\begin{equation}
  g = - \left(\Delta\right)^2\left(dt+ A_1\right)^2 + h_9\,,
 \label{eq:time}
\end{equation}
where $h_9$ stands for a nine dimensional positive definite metric and $A_1$ for a one-form, both defined on a Riemannian
manifold $\cN_9$.
  \item[(ii)] $g(k\,,k)=0$, corresponding to a \emph{lightlike} Killing vector field
\begin{equation}
g = e^{2w(u,x)}[-2du\,dv + H(u,x) (du)^2 +2A_i(u,x)dx^idu + h_8]\,,
 \label{eq:null}
\end{equation}
where $h_8$ stands for an eight dimensional positive definite metric and $A_1 = A_i(u,x)dx^i$ for a one-form, both defined
on a Riemannian manifold $\cN_8$, but also generically depending on the coordinate $u$.
  \item[(iii)] $g(k\,,k)>0$, corresponding to a \emph{spacelike} Killing vector field
\begin{equation}
  g = \left(\Delta\right)^2\left(dz+ A_1\right)^2 + \tilde{g}_9\,,
 \label{eq:space}
\end{equation}
where $\tilde{g}_9$ stands for a nine dimensional Lorentzian metric and $A_1$ for a one-form, both defined on a Lorentzian
manifold $\cM_9$.
\end{itemize}

At this stage, we have only specified the local form of the metric. However, as we will be interested in global issues,
in particular in closed timelike curves, it will be important for us to characterize some global causal features that
we shall require from the set of spacetimes that we shall analyse. In particular, we shall focus on \emph{time-orientable}
spacetimes \cite{HawkingEllis}. A spacetime is called time-orientable if it is possible to define continuously a
division of non-spacelike vectors into two classes, labelled future and past directed \footnote{It is easy to realize that if a spacetime $({\cal M},g)$ is not time-orientable, it always has a double covering space which is time-orientable.}. Thus there is a globally defined timelike vector $\tau$ defining the future direction. A vector $y$ is future directed iff $y^\mu\tau_\mu < 0$ and past directed
iff $y^\mu\tau_\mu >0$ \footnote{Note that this is a weaker condition than the existence of a global time function. The existence of
a time function is equivalent to stable causality. If such a function exists then its gradient is an example of such a vector $\tau_\mu$.}. Obviously for a given time-orientable spacetime, there can be many different choices of such a vector $\tau$. However we show in appendix A that if both $\tau$ and $\tilde{\tau}$ are globally defined timelike vectors which are future directed one with respect to the other, then a timelike vector $y^\mu$ is future (past) directed with respect to $\tau$ iff it is future (past) directed with respect to $\tilde{\tau}$. Therefore the specific choice of $\tau^\mu$ within this class will not affect the classification of vectors into future and past directed.

Given such a time-orientable spacetime, a {\it closed time-like curve} (CTC) is, as its name clearly indicates, a
smooth mapping from the circle to spacetime: $\gamma(\lambda):\,\,S^1\to \cM$, which in local coordinates is given by $x^i(\lambda)$, such that
for all $\lambda\in S^1$, the norm of the tangent vector to $\gamma(\lambda)$ is timelike. There is another requirement on
such a curve, which its name doesn't clearly indicate. The curve must be everywhere future directed \footnote{It can also be everywhere
past directed. The latter is obtained from the future directed one by taking the mapping $\gamma(-\lambda)$.}. This is an important
requirement, because one can always find an everywhere timelike curve from a point to itself going first to the future and then
back to the past in any spacetime.

In this note, we will always consider time-orientable spacetimes which locally admit one of the metrics \eqref{eq:time},
\eqref{eq:null}, \eqref{eq:space}. We will then ask whether they have CTCs, trying to emphasize
the differences that arise depending on whether causal curves propagate in non-compact dimensions or in compact and periodic
ones. Allowing ourselves to consider metrics which locally look like \eqref{eq:time}, \eqref{eq:null} or \eqref{eq:space}, but differ
from them globally, we afterwards analyse the same question for abelian discrete quotients of the latter.

\subsection{Timelike isometry}
\label{sec:time}

Let us start discussing the metrics \eqref{eq:time} possessing a timelike isometry. Being
time-orientable, we can always take the vector $\tau=\partial_t$ to define their future direction. We shall first
discuss the possibility of having CTCs in these spaces when $t\in\bR$ \footnote{The case where $t$ is a compact coordinate can be
thought of a discrete quotient of this space by $\partial_t$ and thus will be discussed in the next subsection.}.

Let $\lambda$ be the affine parameter of a smooth curve $\{t(\lambda),x^i(\lambda)\,\,i=1,\dots ,9\}$. This curve will
be future directed if the condition
\begin{equation}
  \frac{dt}{d\lambda}+A_i(x(\lambda))\frac{dx^i}{d\lambda} > 0 
 \label{eq:futTlike}
\end{equation}
is satisfied at any point $\lambda$. It will be timelike if its tangent vector satisfies at any point
\begin{equation}
  \left\|\frac{d}{d\lambda}\right\|^2 = - \left(\Delta\right)^2\left(\frac{dt}{d\lambda}+ A_i(x(\lambda))\,\frac{dx^i}{d\lambda}\right)^2
  + h_{ij}(x(\lambda))\,\frac{dx^i}{d\lambda}\frac{dx^j}{d\lambda} <0.
 \label{eq:cond1time}
\end{equation}

Since $t$ was assumed to be non-compact, the condition on the closure of the curve gives rise to more severe
constraints, involving the existence of at least one turning point of $t$, i.e. a point $\lambda_*$ where
$\frac{dt}{d\lambda}(\lambda_*)=0$. Thus, a smooth CTC must have a turning point $\lambda_*$ where
\begin{equation}
  \left.A_i(x)\frac{dx^i}{d\lambda}\right|_{\lambda_*} > 0 \;\;\;,\;\;\; [-\Delta^2A_i(x)\,A_j(x) +
  h_{ij}(x)]\,\left.\frac{dx^i}{d\lambda}\frac{dx^j}{d\lambda}\right|_{\lambda_*} < 0~.
 \label{eq:CTCatTP}
\end{equation}
Note that if the curve is constant in the time direction $t(\lambda)=t_0$ then the condition \eqref{eq:CTCatTP}
should be satisfied for all $\lambda$.

In general, there can be analogous requirements associated with the existence of turning points if the curve
$\{t(\lambda)=t_0,\,x^i(\lambda)\}$ propagates along some other non-compact directions. The existence of functions $x(\lambda)$ and a $\lambda_*$ such that \eqref{eq:CTCatTP} is satisfied,
is obviously a necessary but not sufficient condition for the existence of CTCs.

One interesting particular case of the metrics \eqref{eq:time} arises whenever the nine dimensional Riemannian manifold $\cN_9$
has one compact and periodic direction $(\varphi\sim \varphi + 2\pi)$. The integral curves of the vector field
$\partial_\varphi$ would give rise to CTCs whenever
\begin{equation}
  A_\varphi(y^j_0,\,\varphi)> 0 \quad , \quad h_{\varphi\varphi} (y^j_0,\,\varphi) - \Delta^2\,A^2_\varphi (y^j_0,\,\varphi) < 0
  \quad \forall\,y^j_0,\,\varphi ~,
 \label{eq:cond3time}
\end{equation}
where we split $\{x^i(\lambda)\}=\{y^j(\lambda)=y^j_0,\,\varphi(\lambda)\}$. Similar conditions can be derived when
considering more than one compact and periodic dimension.

G\"{o}del-like spaces provide examples of spacetimes having CTCs which fit into this class. In particular, three
dimensional G\"{o}del ($n=1$ in the notation set-up in \cite{Harmark}), corresponds to $\Delta=1$,
$h_{\varphi\varphi}= r^2$ and $A_\varphi=c\,r^2$. The curves $r(\lambda)=r_0 > 1/c$, $x^i(\lambda)=x^i_0$
$i=1,\dots 7$ and $t(\lambda)=t_0$, are examples of CTCs. The same conclusions apply to higher dimensional
G\"{o}del universes \cite{Harmark}.

\subsubsection{Discrete quotients}
\label{sec:dqtime}

Assume the existence of a time-orientable spacetime of the form \eqref{eq:time} having no CTCs. We can consider discrete quotients
of these spacetimes, keeping their metrics locally as \eqref{eq:time}. We shall now analyse which of these abelian quotients
generate CTCs.

As we can only quotient along Killing directions of the metric, let us consider the case where the metric has in addition to $\partial_t$
another isometry $V$. This means that the different tensors building the metric \eqref{eq:time} are left
invariant under its action, i.e. $\cL_V\,\Delta = \cL_V\,A_1 = \cL_V\,h_9=0$. When embedding these metrics in string theory,
there will generically be other matter fields turned on. Therefore, in those cases, it would be assumed that the vector field $V$
leaves the latter also invariant. Under these conditions, it is always possible to work in a coordinate system in which the Killing
vector field is given by $V=\partial_y$. It is in this adapted coordinate system that the full metric \eqref{eq:time} can be written as
\begin{equation}
  g = -\Delta^2\left(dt + A_1\right)^2 + \|\tilde{V}\|^2\left(dy + \tilde{B}_1\right)^2 + \tilde{h}_8~.
 \label{eq:Vtime}
\end{equation}
$\tilde{B}_1$ is a one-form in an eight dimensional Riemannian manifold, whereas the one-form $A_1$
is decomposed as $A_1 = \tilde{A}_1 + A_y\,dy$, so that the norm of the Killing vector is given by
$\|V\|^2= \|\tilde{V}\|^2 - \Delta^2\,A_y^2$, which defines $\|\tilde{V}\|^2$ in \eqref{eq:Vtime}.

The natural discrete quotients of the metric \eqref{eq:Vtime} that we can study can be summarized by
giving the explicit form for the generator of the discrete identification. We write it as:
\begin{equation}
  \xi = \alpha\partial_t + \beta\,V~, \quad \alpha\,,\beta\,\in \bR~,
 \label{eq:xitime}
\end{equation}
where $V=\partial_y$ in \eqref{eq:Vtime}.

It is clear that the spacetime that we get by identifying points along the discrete action generated by
\eqref{eq:xitime} has closed timelike curves whenever $\beta=0$, the reason being that $\|\xi\|< 0$ everywhere, by
assumption. Therefore, the integral curves of the Killing vector $\partial_t$ are both closed and timelike,
providing some particular example. These comments are trivially extended for arbitrary values of
$\{\alpha\,,\beta\}$ whenever the norm $\|\xi\|$ is negative at some point of our manifold. The more interesting
cases are those involving an everywhere spacelike Killing vector $(\|\xi\|>0)$. These are the ones discussed below.

Notice that for $\alpha=0$, and after taking the quotient, $y$ becomes a compact and periodic dimension. This situation
is a particular case of the one outlined in the general discussion above equation \eqref{eq:cond3time}. Integral curves
of the Killing vector $\partial_y$ are not CTCs since we are assuming $\|\xi\|>0$. However there might be other curves which
could be CTCs. In particular, there may exist further periodic dimensions, and by studying curves propagating in these subspaces, one may
discover such curves. For instance, if there is a second periodic dimension $(\varphi)$, there will always be CTCs if the
determinant of the $2\times 2$ metric in the $\{y,\,\varphi\}$ subspace is negative.

We are finally left with the general case $\xi=\partial_t+\beta V$ (where we set $\alpha=1$ without loss of
generality). The norm $\|\xi\|^2=\beta^2\|V\|^2-\Delta^2(1+2\beta A_y)$ is assumed to be positive everywhere. We shall prove
that whenever $V$ is a compact and periodic isometry, i.e. $y \sim y + 2\pi$ is a spatial angle $(\|V\|^2 > 0)$, the
corresponding discrete quotient will \emph{always have CTCs} \footnote{If we would have taken  $V$ to be a compact timelike isometry,
then obviously we create CTCs. The case where $V$ is a null angular variable obviously creates closed null curves. In
the section where we discuss null isometries we consider the creation of CTCs from such orbifolds, where
$\partial_u$ there plays the role of $\partial_t$ here, and $\partial_v$ there plays the role of a null $V$
here.}. Before we give a formal proof, we note that the reason why these CTCs exist is very simple. After the identification,  both
$\partial_y$ and $\partial_t+\beta\partial_y$ are becoming circles. These define a two torus. It is always possible to close a timelike
curve in this two torus by ``waiting long enough'', that is, by going around the two non-trivial circles as many times as necessary,
independently of whether $\beta$ is rational or irrational. This distinction based on the rationality of $\beta$ would certainly play a role
if the curve would be lightlike.

Let us prove the above statements. First, consider a linear transformation in the $\{t\,,y\}$ plane to new coordinates $\{T\,,Y\}$ such that
$\xi=\partial_T$. Thus, after the discrete quotient, $T$ becomes a periodic variable, whereas $Y$ is still a $2\pi$ periodic angle.
As already stressed above, let us focus on curves $\{T(\lambda)\,,Y(\lambda)\,, x^{1..8}(\lambda)=x^{1..8}_0\}$ propagating on the two torus.
The norm of the tangent vector to such a curve is given by
\begin{equation}
  \left\|\frac{d}{d\lambda}\right\|^2 = \|\xi\|^2\left(\frac{dT}{d\lambda} + \frac{\beta\|V\|^2 - \Delta^2\,A_y}{\|\xi\|^2}\,
  \frac{dY}{d\lambda}\right)^2 - \Delta^2 \,\frac{\|\tilde{V}\|^2}{\|\xi\|^2}\,\left(\frac{dY}{d\lambda}\right)^2 ~.
 \label{eq:normtime}
\end{equation}
where $\|\xi\|^2$ is positive by assumption. We are interested in knowing whether it is possible to construct a
closed curve being timelike at the same time.

Let us denote $\alpha\equiv \frac{\beta\|V\|^2 - \Delta^2\,A_y}{\|\xi\|^2}$, which is a constant, independent of
$Y,T,\lambda$. Then we pick two integers $m,n$ such that
\begin{equation}
  \left(\frac{m}{n}+\frac{\alpha}{R}\right) ^2< \frac{\Delta^2 \|\tilde{V}\|^2}{\|\xi\|^4R^2}~,
 \label{eq:nmtlike}
\end{equation}
where $R$ is the radius in the $T$ direction ($T(\lambda=2\pi)=T(\lambda=0)+2\pi R$).
It is easy to verify that the curve $\{T(\lambda)=R\lambda\,,Y(\lambda)=\frac{n}{m}\lambda \}$ is by construction a closed curve
and everywhere timelike. As for its time direction, we note that it is fixed throughout the curve. It is
always future (past) directed if $R+A_y(\frac{n}{m}+\beta R)$ is a positive (negative) number \footnote{Of course
we can always flip the orientation by changing $\{T(\lambda),Y(\lambda)\}$ to $\{T(-\lambda),Y(-\lambda)\}$. Also
note that we can always choose $n,m$ satisfying \eqref{eq:nmtlike} such that the expression
$R+A_y(\frac{n}{m}+\beta R)$ is nonzero.}.

Notice that this construction is very specific to an angle like $V$. Had $V$ not been of this type, the requirement on the
closure of the curve would have put severe constraints on the above construction.

\subsection{Lightlike isometry}
\label{sec:null}

Let us look now at spaces of the form \eqref{eq:null}. Time-orientability implies that there is a globally
well-defined timelike vector field $\tau$ defining the future direction. Let us take this vector to be $\tau =
\frac{e^{-w}}{\sqrt{2}}[\partial_u+(1+H/2)\partial_v]$ which has norm $\|\tau\|^2=-1$. This means that a
vector $y^\mu$ would be \emph{future directed} iff
\begin{equation}
  y^v+y^u(1-H/2)-A_iy^i>0~.
\end{equation}
On the other hand, a vector $y^\mu$ is \emph{timelike} iff
\begin{equation}
  -2y^u[y^v-y^u H/2-A_iy^i]+h_{ij}y^iy^j < 0~.
\end{equation}
Therefore, we see that timelike vectors with $y^u>0$
($y^u<0$) are always future (past) directed, and that future directed timelike curves must have
$y^u=du/d\lambda>0$ everywhere.

In the following discussion we take $u$ to be non-compact. The case in which it is compact can be
viewed as a discrete quotient of this space by $\partial_u$, and therefore will be discussed in the next subsection.

Unlike the previous case of timelike isometry, in the lightlike case it is very easy to argue the non-existence of
closed timelike curves in spacetimes of the form \eqref{eq:null} \footnote{For the special class of metrics with
$w(u,x)$ constant, $A_i=0$ and $h_9=\delta$, this statement was also shown in ~\cite{Sanchez, Mukund}.}. Indeed,
any future directed timelike curve must have $\frac{du}{d\lambda}>0$ everywhere , and any closed curve must have a
turning point in $u$, where $\frac{du}{d\lambda}(\lambda_*)=0$. Obviously the two conditions cannot be satisfied
simultaneously (furthermore, at such a turning point, the norm of the tangent vector to the curve cannot be
negative, as $h_8$ is positive definite. Thus the curve cannot be timelike at that point).

Therefore closed timelike curves in these wave scenarios could only be possible if $u$ is a compact direction.
This can be realized either by considering spacetimes which are already invariant under finite shifts in $u$, i.e.
$w(u,x)$, $H(u,x)$, $A_i(u,x)$ and $h_{ij}(u,x)$ are periodic in $u$ and then one can discretely quotient the
space and make $u$ into an angle, or by considering spacetimes where $u$ is a Killing direction, i.e. all the
above mentioned functions are independent of $u$, and one can then consider a discrete quotient of the spacetime
along $\partial_u$. We analyse this second possibility next.

\subsubsection{Discrete quotients}
\label{sec:dqnull}

As in the discussion of discrete quotients for spacetimes having a timelike isometry, it is convenient
to classify the different discrete quotients that one is going to study by writing the family of Killing vector fields
generating them. In the following, we shall focus on quotients generated by
\begin{equation}
  \xi = \alpha\,\partial_u + \beta\,\partial_v + \gamma\,V~,
 \label{eq:xinull}
\end{equation}
where $V$ stands for an arbitrary Killing vector field of the metric \eqref{eq:null}, that is, it satisfies the
conditions $\cL_V\,w=\cL_V\,A_i=\cL_V\,H = \cL_V h_8 =0$. It is useful to compute the norm of this Killing vector
field $\xi$ :
\begin{equation}
  \|\xi\|^2 = e^{2w}[-2\alpha\beta+H\alpha^2+2\alpha\gamma A_{i}V^i+\gamma^2 h_{ij}V^iV^j]~.
 \label{eq:normnull}
\end{equation}

It is clear from our previous result, that any discrete quotient generated by $\xi$ with $\alpha=0$ has no closed
timelike curves.

Let us set then $\alpha=1$. Whenever $\beta=\gamma=0$, we are effectively dealing with a compact and periodic $u$
coordinate. The integral curves of the Killing vector field $\partial_u$, which are closed by assumption, will be
timelike whenever the fixed point $x^i_0$ in the eight dimensional where they lay is such that $H(x^i_0) < 0$. A
similar result applies to other choices of $\{\beta,\gamma\}$ such that the norm $\|\xi\|^2<0$ at some point of the
manifold. In particular if $\gamma=0$ and $\xi=\partial_u+\beta\partial_v$ then
$\|\xi\|^2=e^{2w(x_0)}[H(x_0)-2\beta]$ and if there is a point $x_0^i$ where $H(x_0)-2\beta<0$ then CTCs will be
created by the quotient.

So we turn to consider the case where $\|\xi\|^2>0$ everywhere (and $\gamma\neq 0$). As before we limit ourselves to
discussing the case where $V$ is a compact and periodic isometry. It is convenient for our purposes to work in an
adapted coordinate system for the eight dimensional metric where $V=\partial_\psi$, $\psi$ standing for a $2\pi$
periodic angle. At the same time, we shall apply a linear transformation in the $\{u\,,v\,,\psi\}$ plane,
\[
  u^\prime = u \quad , \quad v^\prime = v-\beta\,u \quad , \quad \psi^\prime = \psi -\gamma\,u\,
\]
so that the Killing vector field in the new coordinates is $\xi = \partial_{u^\prime}$,
and the metric is given by
\begin{equation}
  \begin{split}
    g  = e^{2\omega (x)}&\left[-2du^\prime\,dv^\prime + \left(H(x)-2\beta\right)\,(du^\prime)^2 + 2 du^\prime\,\left(A_i\,dx^i +
    A_\psi\,\left(d\phi + \gamma\,du^\prime\right)\right)\right. \\
    & \left.  + \|V\|^2\left(d\phi + \gamma\,du^\prime + B_i(x)\,dx^i\right)^2 + l_{ij}dx^i\,dx^j\right]  ~.
  \end{split}
 \label{eq:adnullmetric}
\end{equation}

Let us now construct the following curve: we take $v'(\lambda)=v'_0$ , $x^i(\lambda)=x^i_0$ fixed. Then choose two
integers $m,n$ such that 
$$
  \left|\gamma R-\frac{m}{n}\right|^2 < -R^2\frac{H-2\beta}{\|V\|^2}\;\;,\;\; 
  A_\psi\left(\gamma R-\frac{m}{n}\right)\leq 0~.
$$ 
This is possible iff $H-2\beta<0$ at $x_0^i$. Then we take: $u'(\lambda)=2\pi n R \lambda$ and $\phi(\lambda)=-2\pi m \lambda$ (where $u'$
is periodic with period $2\pi R$). Clearly the curve is closed, and to verify it is timelike, we evaluate:
$$
  \left\|\frac{d}{d\lambda}\right\|^2=e^{2w}(2\pi n R)^2\frac{\|V\|^2}{R^2}\left[R^2\frac{H-2\beta}{\|V\|^2}+\left(\gamma R-\frac{m}{n}\right)^2
  +2\frac{A_\psi R}{\|V\|^2}\left(\gamma R-\frac{m}{n}\right)\right] < 0~.
$$

We conclude that if there is a point $x_0^i$ where $ H(x_0)-2\beta<0$, then the quotient by
$\xi=\partial_u+\beta\partial_v+\gamma \partial_\psi$ for all $\gamma$ creates CTCs \footnote{Actually it can be
proved that for $A_1=0$ and $-H\geq 0$ and superquadratic, also noncompact $V$ Killing vectors would always create
CTCs. This has been shown for $w=0$ in \cite{Sanchez} and can be trivially extended to the case where $w\neq 0$.
}.

The above existence proof uncovers all plane wave metrics propagating in $\bR^8$, which after taking a discrete quotient
by $\xi=\partial_u + \beta\partial_v + \gamma\,V$ whose norms equal one, $\|\xi\|^2=1$, give rise to spacetimes having CTCs
and being T-dual to G\"{o}del-like universes \cite{Horava,Harmark,Mukund}. It generalises to arbitrary pp-wave metrics propagating
in generic curved eight dimensional euclidean manifolds, having at least one compact and periodic dimension.

\subsection{Spacelike isometry}
\label{sec:space}

Let us finally consider metrics of the form \eqref{eq:space}. Given a curve $\{z(\lambda),\,x^i(\lambda)\}$, the norm of
its tangent vector at a point $\lambda$ is given by
\begin{equation}
  \left\|\frac{d}{d\lambda}\right\|^2 = (\Delta)^2\left(\frac{dz}{d\lambda} + A_i(x(\lambda))\,\frac{dx^i}{d\lambda}
  \right)^2 + \tilde{g}_{ij}\frac{dx^i}{d\lambda}\,\frac{dx^j}{d\lambda}~.
\label{eq:normsp}\end{equation}

This curve is a CTC iff the above norm is negative everywhere, the curve is closed and it is future directed.

Let us now consider the 9-dimensional spacetime with metric $\tilde{g}_9$. As the 10-dimensional spacetime is time-orientable, this 9-dimensional one must also be, 
and if $\tau=\tau^z\partial_z+\tau^i\partial_i$ was a future directed vector in the 10-dimensions, it is easy to see that $\tau^i\partial_i$ is a timelike 
vector in the 9-dimensional spacetime, defining its future direction.

We now show that the curve $\{x^i(\lambda)\}$ is a CTC in the 9-dimensional spacetime with metric $\tilde{g}_9$. It is straightforward to see that as a 
projection of the 10-dimensional closed curve, it must also be closed, and from \eqref{eq:normsp} it is clear that it must also be timelike. It only remains 
to show that it is future directed with respect to $\tau^i\partial_i$. This fact is proven in appendix B.

Thus, we find that a necessary condition for the existence of CTCs in these scenarios, is the existence of CTCs in
the nine dimensional metric $\tilde{g}_9$.

The discussion based on non-propagating curves in $z$ effectively reduces to discussions presented in previous
sections, depending on the amount and kind of isometry present at sections of constant $z$. We shall therefore skip this
point and concentrate on the more interesting compact $z$ isometry.

\subsubsection{Discrete quotients}
\label{sec:dqspace}

Let us proceed as we did for the previous discrete quotient discussions. In this case, we shall study discrete
quotient manifolds obtained by identifying points under the discrete action generated by
\begin{equation}
  \xi = \alpha\,\partial_z + \beta\,V~.
 \label{eq:xispace}
\end{equation}

If we set $\beta=0$, we end up with the standard Kaluza--Klein ansatz metric, where now $z$ is compact and periodic.
We shall next prove that in this case, the previous necessary condition for existence of closed timelike
curves is also sufficient. Indeed, if we assume the existence of a closed timelike curve in $\tilde{g}_9$, we can always construct a closed
timelike curve in ten dimensions. Denote such a nine dimensional closed curve by $\{x^i(\lambda)\}$, $\lambda\in\,[0,2\pi]$,
and introduce a real number $\kappa\in\bR$ satisfying\footnote{Such a number must exist as $x^m(\lambda)$ are
all smooth functions from a compact domain to a compact domain.} :
\begin{equation*}
  \Big|\tilde{g}_{mn}\frac{dx^m}{d\lambda}\frac{dx^n}{d\lambda}\,g_{zz}^{-1}
  \Big| \geq \kappa^2 \quad \forall\,\lambda\,.
\end{equation*}
The crucial steps of the proof consist of showing that we can always construct a ten dimensional closed timelike
curve by requiring
\begin{equation}
  \frac{dz}{d\lambda} = -A_m(x(\lambda))\,\frac{dx^m}{d\lambda} + \epsilon ~,
 \label{eq:10ctcspace}
\end{equation}
and demanding $\epsilon^2 < \kappa^2$. First we want to show that the curve resulting
from integration of the differential equation \eqref{eq:10ctcspace} can be made closed. Indeed, we want to know whether
there is a real number $\epsilon$ , and two integers $k,n\in\bZ$ for which
\begin{equation*}
  z(2\pi k) = z(0) +2\pi R\,n
\end{equation*}
where $2\pi$ is the period in the affine parameter of the nine dimensional closed curve $x^m(\lambda)$, and $R$ is the radius in the $z$ direction.
The answer is clear, choose
\begin{equation*}
  \epsilon = \frac{1}{2\pi k}\int^{2\pi k}_{0} A_m(x(\lambda))\,
  \frac{dx^m}{d\lambda}\,d\lambda + R\frac{n}{k} \equiv a + b\,\frac{n}{k}\,,
\end{equation*}
where $a\,,b\in\bR$ \footnote{Note that $a$ is independent of $k$ as the curve $x^m(\lambda)$ is periodic.
In fact $a=\frac{1}{2\pi}\int_0^{2\pi} A_m(x(\lambda))\,\frac{dx^m}{d\lambda}\,d\lambda$.}.
The second step consists of showing that the closed curve is timelike. This is equivalent
to showing that one can always choose two integers $\{n,k\}$ such that
\[
  \left|a^\prime + b^\prime\,\frac{n}{k}\right| < 1\,, \quad a^\prime\,,b^\prime\in\bR
\]
which is a condition that is derived from the choice of $\epsilon$ and the condition $\epsilon^2 <
\kappa^2$. That one can indeed find these integers can be seen just by choosing
\begin{equation*}
  n = \text{Int}\Big(\frac{a^\prime}{b^\prime}\,k\Big) \pm 1 \quad
  , \quad k = \text{Int}(b^\prime) \pm 1 \,.
\end{equation*}

The previous discussion was focusing on the closure and timelike character of the curve, but as we stressed in the introduction of
section \ref{sec:ctcsymm}, a further important requirement comes from being a future directed curve. Thus, the last step of the proof consists
in showing that the ten dimensional curve that we constructed is indeed everywhere future directed. Assume the nine dimensional curve $x^m(\lambda)$
is future directed with respect to $\tilde{\tau}$, so that this curve is indeed a good CTC in nine dimensions. We can always build a ten dimensional
globally defined timelike vector field $\tau$ out of $\tilde{\tau}$. Indeed, we can just take $\tau= -(A_j \tau^j)\partial_z + \tilde{\tau}$.
Since we proved in appendix \ref{sec:appA} that future directness is independent of the choice of the representative in the class
of mutually future directed timelike vectors, we only have to prove that our ten dimensional curve is future directed with respect to this
particular choice of $\tau$. This last statement is trivial, if it is also satisfied in nine dimensions, as we assumed.

This completes the proof that the ten dimensional metric \eqref{eq:space} quotiented by $\partial_z$ has closed
timelike curves if and only if the nine dimensional one $\tilde{g}_9$ does. This statement will have important
consequences for the behaviour of closed timelike curves under everywhere spacelike T-duality transformations, but
we shall postpone this discussion until section \ref{sec:ctcdual}.

Let us now consider the general scenario, in which $\beta\neq 0$, and for convenience set $\alpha=1$. If $V$
is timelike or null, then this is one of the cases we already considered in sections 2.1 or 2.2. We concentrate,
as before, on the case where $V$ is a spacelike compact rotational isometry.

By assumption, the description of such an spacetime can always be given in terms of the following metric description :
\begin{equation*}
  g = (\Delta)^2\,\left(dz + A_i\,dx^i + A_\psi\,d\psi\right)^2 + \|\tilde{V}\|^2\,\left(d\psi + B_i\,dx^i\right)^2 + \tilde{g}_8~,
\end{equation*}
where $V=\partial_\psi$, and $\|V\|^2 = \|\tilde{V}\|^2 + \Delta^2\,A_\psi^2$.
By a linear transformation, $Z=z,\, \phi=\psi-\beta z$, the Killing vector becomes $\xi=\partial_Z$ and the final metric
can be written in the form
\begin{equation}
  g = \|\xi\|^2\,\left(dZ+a_\phi d\phi+ a_1\right)^2 + Q\left(d\phi+P_1\right)^2 +h_8~,
\end{equation}
where $a_1,p_1,h_8$ are defined on the 8-dimensional manifold transverse to $Z,\phi$ and the number $Q$ is given in terms of the geometric
data defined above by
\begin{equation*}
  Q = \frac{\Delta^2\,\|\tilde{V}\|^2}{\|\xi\|^2} = \frac{\Delta^2\left(\|V\|^2 - \Delta^2\,A_\psi^2\right)}{\|\xi\|^2}~.
\end{equation*}

By our previous discussions, we know that whenever $\|\xi\|^2<0$ we will have CTCs, which corresponds to the condition
\begin{equation*}
  \|V\|^2 < -\frac{\Delta^2}{\beta^2}\,\left(1+2\beta\,A_\psi\right)~.
\end{equation*}
If $\|\xi\|^2>0$, but $Q<0$, there are also CTCs. Both conditions translate into
\begin{equation*}
  \Delta^2\,A_\psi^2 > \|V\|^2 > -\frac{\Delta^2}{\beta^2}\,\left(1+2\beta\,A_\psi\right)~.
\end{equation*}
Finally, if both $\|\xi\|^2,\,Q > 0$, by trivial extension of our arguments, there will exist CTCs iff the 8-dimensional
metric $h_8$ has CTCs.

\subsection{On closed timelike curves and U-duality}
\label{sec:ctcdual}

In the previous section, we have discussed under which circumstances closed timelike curves exist in certain
families of metrics. It is natural to ask whether there exists some relation between different metrics having
closed timelike curves, and even more intriguing, whether metrics having them are related to metrics free of these
causality violating curves under certain duality transformations. In particular, it is natural to ask whether the
notion of existence of closed timelike curves is a U-duality invariant notion. Before entering into a more
detailed discussion, we would like to emphasize the regime of validity in which we will be working. The whole
notion of a closed timelike curve is based on geometrical (and metric) grounds, and as such, any discussion of its
existence will entirely rely on a classical description of a given spacetime metric. Furthermore, when dealing
with T-duality in string theory, it is important to distinguish how this conjectured duality manifests in the
different regimes of the theory. Since we are forced to work in the classical gravitational limit, T-duality is
manifested through the Buscher rules \cite{buscher} mapping on-shell configurations belonging to ``different
supergravity theories''. These transformation laws are only valid to one loop in $\alpha^\prime$. Furthermore,
they do not take possible physical effects associated with the winding sector of the theory into account.
Therefore, all we have to say below is concerned with the purely classical general relativistic approach to
T-duality. But even in this limit, there are non-trivial statements to be made due to the interplay between the
metric and the NS-NS two form.

Let us assume that we have a classical solution to some type II supergravity theory
which is reliably described in this regime\footnote{The dilaton does not blow up, and the curvature $\cR$ is small in
string units, $\cR\cdot\alpha^\prime\ll 1$.}. Concerning S-duality, it is manifest, at least at the level of type
IIB supergravity, that the existence of closed timelike curves is an {\it S-duality invariant notion}. The reason
is that the metric of the original configuration and its S-dual description are conformally related. Thus, the
causal structure of both metrics is the same, and in particular, the existence or non-existence of closed timelike
curves is preserved under this particular kind of transformations. Notice that this is a general statement which
does not depend on the kind of spacetime we are studying.

On the other hand, a priori, it would seem that such a property is not
shared by T-duality. By looking at the transformations mapping the supergravity multiplets between type IIA and
IIB supergravity, one observes that the metric and NS-NS two form are related to each other, so that it is far
from obvious that such a property would be preserved. Even though it may be counterintuitive, there is an argument
giving some clues that this might be correct. Any on-shell spacetime supergravity configuration to which we want
to apply a (spacelike) T-duality transformation\footnote{The possibility of studying T-duality along a timelike
direction can also be considered, and one can work out some generalization of the T-duality rules.
We will not discuss this possibility in this note.} has to be invariant under a spacelike $z$ translation\footnote{From
now on, the $z$ direction will stand for the direction along which we study T-duality.}. In particular, this means
that the original metric configuration can always be written in a Kaluza--Klein ansatz form :
\begin{equation}
  g = \tilde{g}_{mn}\,dx^m\,dx^n + g_{zz}\left(dz+A_m dx^m\right)^2\,.
 \label{eq:kkansatz}
\end{equation}
Therefore, the metric \eqref{eq:kkansatz} is of the form \eqref{eq:space}, and we can use all the results that we
obtained in subsection \ref{sec:dqspace} for a periodic spacelike isometry $\partial_z$.

Physically, it is clear that in the limit in which the Kaluza--Klein observer description is reliable (nine dimensional metric
$\tilde{g}$), the notion of existence of closed timelike curves will be T-duality invariant, since for that
observer, only the nine dimensional metric that he/she measures is relevant, and that metric
$(\tilde{g})$ is T-duality invariant. We want to show that, actually, the notion of existence or non-existence of
closed timelike curves is a ten dimensional {\it T-duality invariant notion}.

In order to prove this statement, let us first write the T-dual metric of a given metric \eqref{eq:kkansatz}
\begin{equation}
  g^\prime = \tilde{g}_{mn}\,dx^m\,dx^n + \frac{1}{g_{zz}}
  \left(dz-B_{mz} dx^m\right)^2\,,
 \label{eq:tkkansatz}
\end{equation}
where $B_{mz}$ stands for the $i_{\partial_z}B_2$ components of the original NS-NS two form. Notice that
\eqref{eq:tkkansatz} is again of the form \eqref{eq:space}.

We already know that both 10-dimensional metrics would have CTCs iff the 9-dimensional reduced one $\tilde{g}$ does, a statement
that relies on appendix \ref{sec:appB}. Since for both T-dual metrics, this Kaluza--Klein reduced metric is the same, we conclude
that the existence of CTCs is a T-duality invariant notion \footnote{In \cite{Carlos2}, it was claimed that
in certain cases the T-dual closed timelike curves could be "resolved" in the sense that they could change their
nature from being purely topological to purely geometrical. The attitude in that paper is that all closed timelike
curves introduced by discrete identifications are not intrinsic to Einstein field equations, since the latter are
local equations, and locally our quotients are not distinguishable from the original spacetime that we started
from. Thus, one might be tempted to conclude that from a purely supergravity point of view the latter distinction
is important. In the present paper, we prefer to take a more stringy perspective, and even though, our analysis
only applies to the supergravity regime, we do believe the physics of our original spacetime and the one
after discrete identifications are manifestly different. Thus, we do want to claim there are still closed
timelike curves in the T-dual description.}

Knowing this fact about T-duality and CTCs, it is natural to ask whether there is any relation between different
scenarios discussed in section \ref{sec:ctcsymm} having CTCs. As it is already known in the literature, there
is a close relationship between certain compactified pp-waves and G\"{o}del-like universes. We are in a position
to generalize this connection to a wider set of configurations.

Consider a discrete quotient of a type II configuration having a lightlike
isometry, so it is described by \eqref{eq:null}, with a vanishing NS-NS two-form, and some non-trivial dilaton profile
\footnote{We can certainly include arbitrary RR form potentials, but they will not modify our argument, so we shall
not include them in the discussion to keep it free of useless technicalities.}, which has CTCs.
Thus, the generator of the identifications is given by
\[
\xi=\partial_u + \beta\,\partial_v + \gamma\,V~,
\]
where $V$ stands for a compact and periodic isometry. We can introduce an adapted coordinate system
\begin{equation}
  \begin{aligned}[m]
    \vec{x} & = e^{\gamma\,u\,V}\,\vec{y} \\
    v &= v^\prime + \beta\,u~,
  \end{aligned}
 \label{eq:uadapted}
\end{equation}
in which $\xi=\partial_u$. The metric \eqref{eq:null} written in coordinates \eqref{eq:uadapted} is given by
\begin{equation}
  g = e^{2\omega}\,[-2du\,dv^\prime + g_{ij}dy^i\,dy^j + 2(A_{(1)}+ \gamma\,V_{(1)})\,du] + \|\xi\|^2\,du^2~,
 \label{eq:adaptednull}
\end{equation}
where $A_{(1)}+ \gamma\,V_{(1)} = A_i(\vec{y})\,dy^i + \gamma\,g_{ij}(\vec{y})V^j(\vec{y})\,dy^i$.
Applying the T-duality rules appearing in \cite{buscher} along the spacelike direction $\partial_u$, one obtains the type II T-dual configuration
\begin{equation}
  \begin{aligned}[m]
    g & = -e^{4\omega}\|\xi\|^{-2}\left(dv - (A_{(1)}+ \gamma\,V_{(1)})\right)^2 + e^{2\omega}\,g_{ij}dy^i\,dy^j +
    \|\xi\|^{-2}\,du^2 \\
    B_{(2)} & = e^{2\omega}\,\|\xi\|^{-2}\,du\wedge \left(A_{(1)}+ \gamma\,V_{(1)}-dv\right) \\
    e^{2\phi^\prime} &= e^{2\phi}\, \|\xi\|^{-2}~.
  \end{aligned}
 \label{eq:tdualumetric}
\end{equation}
Therefore, we can state that the T-dual configuration of discrete quotients of pp-wave backgrounds \eqref{eq:null} with vanishing
NS-NS two-form and Killing vector \eqref{eq:xinull}, giving rise to CTCs after the discrete identification, is a metric of the type
\eqref{eq:time}. This conclusion uncovers many particular examples that have already been reported in the literature and points
out that the crucial property behind these relations are the symmetries of the spacetimes involved in the discussion.

\section{Supersymmetric configurations and CTCs}
\label{sec:juanliat}

The extension of the formalism developed in the previous section to supersymmetric configurations
is straightforward. The only relevant question that remains open is if supersymmetry is preserved under discrete quotients of the initial spacetime configuration, and if so, how many such supersymmetries are preserved.

The answer to this question can be non-trivial in general. To begin with, the quotient manifold  $\cM/\Gamma$
may not even allow a spin structure, so that standard fermions do not exist on it \footnote{This does not
exclude the possibility of having ``charged'' fermions in some $\text{spin}^c$ structure, for example.}.
Since it is not the purpose of this paper to investigate this issue, we shall assume that $\cM/\Gamma$
has spin structure. In that case, given a background with Killing spinors $\varepsilon$, the local criterium
for the existence of supersymmetry in $\cM/\Gamma$ is given by
\begin{equation}
  \cL_\xi \varepsilon = \xi^M\nabla_M\varepsilon-\frac{1}{4}\nabla_{[M}\xi_{N]}\Gamma^{MN}\varepsilon =0~,
 \label{eq:susyquotient}
\end{equation}
which selects the subset of the original Killing spinors left invariant under the action of the generator
$\xi$ of the discrete identification. We shall use this approach in the following subsections.

\subsection{Timelike isometries in susy configurations}
\label{sec:susytime}

As an example of closed timelike curves in a supersymmetric realization of the metric family \eqref{eq:time}, we
shall consider a discrete quotient of the well-known supergravity configuration describing a bunch of parallel
D3-branes in Minkowski spacetime. We would like to stress that the particular example discussed below is just the
type IIB adaptation of the scenarios discussed in ~\cite{jose2,jose3} in an M-theory context and which were
referred in these references to as ``exotic reductions''.

The idea is very simple. Consider the classical type IIB configuration describing D3-branes located at the origin,
whose metric\footnote{The full type IIB configuration involves a constant dilaton and a non-trivial
RR five-form field strength, but we shall not be concerned with their explicit form in the following discussion.}
is given by
\begin{equation*}
  g = f^{-1/2}(r)\, ds^2(\bE^{1,3}) + f^{1/2}(r)\, ds^2(\bE^6) ~, \quad f(r)= 1 + \frac{|Q|}{r^4}~,
\end{equation*}
and study its quotient under a discrete identification generated by the Killing vector field
\begin{equation}
  \xi = \tau_\parallel + \rho_\perp = \mu\,\partial_t + \theta_1\,R_{45} + \theta_2\,R_{67} + \theta_3\,R_{89}~,
\end{equation}
where $R_{ij}$ stands for a rotation in the $ij$-plane.

The first question one needs to clarify is whether the norm of the above Killing vector can be made spacelike
everywhere, by a convenient choice of the set of parameters $\{\mu,\,\theta_i\}$. Notice that the norm
of the transverse rotation at infinity satisfies some bounds
\begin{equation*}
  r^2M^2 \geq \|\rho_\perp\|_\infty^2 \geq r^2m^2~,
\end{equation*}
where $m$ can be made nonzero if $\theta_i\neq 0$ $\forall\,i$. Therefore, the norm of the Killing vector satisfies
\begin{equation*}
  \|\xi\|^2 = -\mu^2\,f^{-1/2} + f^{1/2}\|\rho_\perp\|_\infty^2 \geq -\mu^2\,f^{-1/2} + f^{1/2}r^2m^2~,
\end{equation*}
since $f(r)$ is an everywhere positive function. The right hand side defines a function of the radial coordinate $F(r)$.
This function is minimized at the critical radius
\begin{equation}
  r_0^2m^2 = \frac{|Q|\,\mu^2}{r_0^4 + |Q|}~,
 \label{eq:critical}
\end{equation}
as can be checked by evaluating the second derivative $F''(r_0)$. Requiring the value of the function at the critical
radius to be positive, to ensure the spacelike character of the Killing vector $\xi$ everywhere, gives the extra
condition $r_0^4 \leq |Q|$, which can be translated in terms of the $\mu$ parameter appearing in the determination
of the critical radius \eqref{eq:critical}
\begin{equation}
  \mu^2 \leq 2m^2|Q|^{1/2}~.
 \label{eq:upbound}
\end{equation}
To sum up, it is possible to construct a discrete quotient of the standard D3-brane configuration involving timelike
translations and transverse rotations leaving no directions perpendicular to the brane invariant $(\theta_i\neq 0\,
\forall\,i)$, such that the norm of the full Killing vector is spacelike everywhere. This is achieved by fine tuning
the otherwise free parameter $\mu$, which is constrained to satisfy the upper bound \eqref{eq:upbound}.

Since $\rho_\perp$ corresponds to a compact and periodic isometry, the above example fits into our general discussion
of discrete quotients having a timelike isometry. As such, we conclude it has CTC's. The same conclusion applies
to all ``exotic reductions'' considered in \cite{jose2,jose3}. It is important to stress that the above conclusion
is independent of whether the quotient preserves supersymmetry or not, that is, it is independent of the choice
of the rotation parameters $\{\theta_i\}$ as far as all of them are non-vanishing. For a discussion on the supersymmetry
preserved by these quotients, see \cite{jose2}.

\subsection{Null isometries of the pp-wave type}
\label{sec:susynull}

We shall now discuss some examples of metrics having null isometries in a supersymmetric context. We shall focus
on the type IIB supergravity configurations studied in \cite{juanpp}, even though there were more general
backgrounds both in type IIB and IIA discussed in \cite{juanppfup}. Let us quickly review the solutions presented
in \cite{juanpp}.

The general (2,2) supersymmetric pp-wave background can be parameterized in the form:
\begin{equation}
  \begin{aligned}[m]
    g &= -2dx^+dx^- -32(|dW|^2+|V|^2)(dx^+)^2+2g_{\mu\bar{\nu}}dz^{\mu} d{\bz}^{\bnu}~, \\
    F_5 &= dx^+\wedge \varphi_4~, \\
    \varphi_{\mu\nu} &=\nabla_\mu\nabla_\nu W ~~,~~ \varphi_{\bmu\bnu}=\nabla_{\bmu}\nabla_{\bnu}\overline{W} ~~,~~
    \varphi_{\bmu\nu}=\nabla_{\bmu}\nabla_{\nu}U~,
  \end{aligned}
 \label{eq:liatwo}
\end{equation}
where $W$ is a holomorphic function, and $V^\mu$ a holomorphic Killing vector coming from a real Killing potential
$U$ such that $\nabla_\mu V^\mu=0$ and $\partial_\nu[V^\mu\nabla_\mu W]=0$. $\varphi_2$ parameterizes the four-form $\varphi_4$ in $F_5$ as explained in
\cite{juanpp} :
\[
  \varphi_4=\frac{1}{4!}[\varphi_{\mu\nu}g^{\nu\bar{\nu}}\epsilon_{\overline{\rho\sigma\lambda\nu}}dz^{\mu}\wedge d\bar{z}^{\bar{\rho}}\wedge
  d\bar{z}^{\bar{\sigma}}\wedge d\bar{z}^{\bar{\lambda}} +
  \frac{1}{2}\varphi_{\mu\bar{\nu}}g_{\rho\bar{\rho}}dz^{\mu}\wedge d\bar{z}^{\bar{\nu}}\wedge dz^{\rho}\wedge d\bar{z}^{\bar{\rho}}+c.c.]~.
\]
The Killing spinors are parameterized by two complex parameters.

On the other hand, the general (1,1) supersymmetric pp-wave background can be parameterized in the form:
\begin{equation}
  \begin{aligned}[m]
    g &= -2dx^+dx^- -32(|dU|^2)(dx^+)^2+2g_{\mu\bar{\nu}}dz^{\mu} d{\bz}^{\bnu} \nonumber\\ F_5 &= dx^+\wedge \varphi_4 ~,\\
    \varphi_{\mu\nu} &=\nabla_\mu\nabla_\nu U ~~,~~ \varphi_{\bmu\bnu}=\nabla_{\bmu}\nabla_{\bnu}U ~~,~~
    \varphi_{\bmu\nu}=\nabla_{\bmu}\nabla_{\nu}U~,
  \end{aligned}
 \label{eq:liatone}
\end{equation}
where $U$ is a real harmonic function. The Killing spinors are parameterized by one complex parameter, and $\varphi_4$ is given in
terms of $\varphi_2$ as explained above.

These solutions are free of closed timelike curves, but we would like to study whether discrete quotients of them can generate such curves
while preserving supersymmetry. By our general arguments in section \ref{sec:null}, many of these discrete quotients will indeed have CTCs, so
we are left to determine which of these quotients preserve some amount of supersymmetry. This is analysed in detail in appendix \ref{sec:appC}.
The main conclusion out of this analysis is that any Killing vector of the form
\begin{equation}
  \xi = a\partial_+ + b\partial_- + \tilde{V}\,, \quad a,\,b\in\bR
 \label{eq:susykill}
\end{equation}
where $\tilde{V}$ is an holomorphic Killing vector of the transverse eight dimensional metric such that it commutes
with $V$, i.e. $[V\,,\tilde{V}]=0$, and preserves $\nabla\,W$, i.e. $\cL_{\tilde{V}}\nabla_\mu W =0$, preserves all the
(2,2) supersymmetries of the original background. For completeness, we provide its norm
\begin{equation}
  \|\xi\|^2 = -32a^2\left(|dW|^2 + |V|^2\right) + 2|\tilde{V}|^2- 2ab\,.
 \label{eq:norm}
\end{equation}

According to our previous discussion in section \ref{sec:ctcsymm}, we can immediately conclude that for $a=0$ we get no CTCs
in the discrete quotient spacetime. On the other hand, for $a\neq 0$, any discrete quotient generated by $\xi$ with $\tilde{V}$
being a compact and periodic isometry can give rise to CTCs. We shall study a couple of examples where this feature will be
shown explicitly.

\subsubsection{Flat transverse space}
\label{sec:flat}

The simplest example we could consider is a pp-wave propagating in a flat eight dimensional transverse space
\begin{equation}
  \begin{aligned}[m]
    g &= -2dx^+dx^- -32(|dW|^2+|V|^2)(dx^+)^2+2\delta_{\mu\bar{\nu}}dz^{\mu} d{\bz}^{\bnu}~, \\
    \varphi_{\mu\nu} &= \partial_\mu\partial_\nu W ~~,~~ \varphi_{\bmu\bnu}=\partial_{\bmu}\partial_{\bnu}\bar{W} ~~,~~
    \varphi_{\bmu\nu}=\partial_{\bmu}\partial_{\nu}U~.
  \end{aligned}
\end{equation}
Out of the symmetries of this spacetime, we shall only consider those preserving its complex structure, thus
\[
  V= i(P^\mu+R^\mu_{\;\nu}z^\nu)\partial_\mu\,+c.c.\,\, , \,\, U=P_\mu z^{\mu}+R_{\mu\nu}z^\mu z^\nu+c.c.~,
\]
and $W$ is any holomorphic function.

According to our general analysis in the previous section, the generator of the discrete quotient
\[
  \xi = a\partial_++b\partial_-+\tilde{V}\;\;,\;\; \tilde{V}=i(\tilde{P}^\mu+\tilde{R}^\mu_{\;\nu}z^\nu)\partial_\mu\,+\,c.c.~,
\]
will preserve all the original supersymmetries if $\tilde{V}$ is an holomorphic Killing vector field satisfying
$[V,\tilde{V}]=0$ and ${\cal L}_{\tilde{V}}\nabla\,W=0$. These conditions imply that
\[
  [R,\tilde{R}]=0\;\; , \;\; P^{\mu}\tilde{R}^\nu_{\;\mu}=\tilde{P}^{\mu}R^{\nu}_{\;\mu}\;\; , \;\;
  (\tilde{P}^\mu+\tilde{R}^{\mu}_{\;\nu}z^{\nu})\partial_{\mu}W=\text{const}~.
\]

We introduce new coordinates
\begin{equation}
  \begin{aligned}[m]
    (x^-)^\prime &= x^- -(b/a)x^+~,\\
    y &= e^{-\tilde{V}\,x^+}\cdot z~,
  \end{aligned}
\end{equation}
in which $\xi = \partial_{x^+}$, by construction. The metric is given by
\begin{equation}
 g =-2dx^+dx^{-'}+\|\xi\|^2(dx^+)^2+2[\tilde{V}_{\mu}dy^{\mu}+\tilde{V}_{\bmu}d\bar{y}^{\bmu}]dx^+
 +2\,\delta_{\mu\bar{\nu}}\,dy^{\mu} d\bar{y}^{\bar{\nu}}~,
\end{equation}
where
\[
  \|\xi\|^2 = -32a^2\left(|dW|^2 + |V|^2\right) + 2|\tilde{V}|^2- 2ab~.
\]

The identification $x^+\sim x^+ + 2\pi R$ generates CTCs, and since we know these are preserved under a T-duality transformation,
it is just more convenient for us to look for them in the T-dual picture. Taking the particular case $W=0$, $\tilde{V}=4aV=
iR^\mu_{\;\nu}y^\nu\partial_{\mu}\,+c.c.$ and $2ab=-1$, one finds that $\|\xi\|^2=1$, i.e. $\xi$ is everywhere spacelike.
After T-dualizing the metric along $\xi$, we get the family of G\"{o}del-like solutions related to compactified plane waves,
mentioned in \cite{Horava,Harmark,Mukund}. Their metrics are given by
\begin{equation*}
 \tilde{g}=-[dx^{-'}-(iR^{\mu}_{\;\nu} y^{\nu} d\bar{y}^{\bmu}+c.c)]^2+(dx^+)^2 + 2\,\delta_{\mu\bar{\nu}}\,dy^{\mu} d\bar{y}^{\bar{\nu}}~,
\end{equation*}
which are known to have CTCs.

\subsubsection{Eguchi--Hanson transverse space}
\label{sec:eh}

Another very interesting metric to look at, which possesses some supersymmetry is a pp-wave whose 8-dimensional
transverse space $z^M,\overline{z^M}$, $M=1..4$ is the direct product of an Eguchi--Hanson space on
$z^{\mu},\overline{z^{\mu}}$, $\mu=1,2$ and a $\bC^2$ space $z^{i},\overline{z^i}$, $i=3,4$.

The Eguchi--Hanson metric is given, in the above complex coordinates, in terms of the K\"{a}hler potential:
\begin{equation*}
  K = \sqrt{\rho^4+a^4}-a^2\ln \left(\frac{a^2}{\rho^2}+\sqrt{1+\frac{a^4}{\rho^4}}\right) = r^2 +\frac{a^2}{2}\ln \left(
  \frac{r^2-a^2}{r^2+a^2}\right)~,
\end{equation*}
where $\rho^2\equiv (z^1)^2+(z^2)^2$ and $r^4\equiv \rho^4+a^4$. Its components are explicitly given by
\begin{equation}
  g_{\mu\bar{\nu}} = \sqrt{1+\frac{a^4}{\rho^4}}\delta_{\mu\bar{\nu}}-\frac{a^4 z_\mu\overline{z_\nu}}{\rho^6\sqrt{1+\frac{a^4}{\rho^4} }} =
  \sqrt{1+\frac{a^4}{\rho^4}}\left(\delta_{\mu\bar{\nu}}-\frac{a^4 z_\mu\overline{z_\nu}}{\rho^2(\rho^4+a^4 )}\right)~.
 \label{eq:EHmet}
\end{equation}
Thus, the full transverse space where the pp-wave propagates is given by
\[
  h_8 = 2[g_{\mu\bar{\nu}}z^{\mu}\overline{z^{\nu}}+\delta_{i\bar{j}}dz^id\overline{z^j}] = g_{\text{EH}} + g_{\bC^2}~.
\]
The latter has the following traceless holomorphic Killing vectors: $V^{M}=iC^M_{~N}z^N$ where $C^M_{~N}$ is a constant
hermitian traceless 4x4 matrix which is block diagonal\footnote{Note that $V^M$ cannot have a constant piece in the 1,2 directions, as it must
respect the $Z_2$ symmetry $z^{\mu}\to-z^{\mu}$ of the Eguchi--Hanson space.}.

Although it is possible to work with a general such matrix (having seven complex parameters), we choose to
simplify things and work with the diagonal matrix $C=\mu \cdot diag(1,1,-1,-1)$, $\mu$ a real parameter. For this
choice, the norm of the Killing vector is $\|V\|^2= \mu^2(|z^3|^2+|z^4|^2 +\frac{\rho^4}{r^2})$. All in all,
we shall be considering the following particular (2,2) supersymmetric pp-wave solution \footnote{We could have also added any holomorphic
function $W(z^M)$ to $g_{++}$ and the corresponding 5-form fluxes, such that $V^{M}\partial_MW=const$} :
\begin{equation}
  \begin{aligned}[m]
    g  &= -2dx^+dx^- -32|V|^2(dx^+)^2 + g_{\text{EH}} + g_{\bC^2}~, \quad \|V\|^2 = \mu^2(|z^3|^2+|z^4|^2 +\frac{\rho^4}{r^2})~, \\
    \varphi_{1\bar{1}2\bar{2}} &= \varphi_{3\bar{3}4\bar{4}}=\mu\;\;,\;\;\varphi_{\mu\bnu i\bar{j}}=\frac{\mu}{2}
    \delta_{i\bar{j}}[\partial_{\mu}g_{\bnu\lambda}]z^\lambda~.
  \end{aligned}
\end{equation}
We know there are no CTCs in the above pp-wave configuration, so we shall focus on discrete quotients of it generated
by the Killing vector field
\[
  \xi = \partial_+ +a\partial_- + (iD^M_{~N}z^N\partial_M +c.c)~,
\]
where $D^M_{~N}$ stands for a 4x4 constant block diagonal hermitian and traceless matrix. Instead of studying the most general scenario, we
specialize to the simpler case where $D=\nu\cdot diag(1,1,-1,-1)$, $\nu$ real.

It is convenient to introduce a new set of coordinates in which $\xi = \partial_+$. The latter is given by
\begin{equation}
  \begin{aligned}[m]
    x^{-'} &= x^- - a\,x^+~,\\
    \omega^{1,2} &= e^{-i\nu x^+}\,z^{1,2}\;\;\;,\;\;\;\omega^{3,4} = e^{i\nu x^+}z^{3,4}~.
  \end{aligned}
\end{equation}
The metric, when written in terms of this adapted coordinate system, becomes
\begin{multline}
  g =-2dx^+dx^{'-} + \|\xi\|^2\,(dx^+)^2  + g_{\text{EH}} + g_{\bC^2} \\
  - i\nu \left[(\omega^\mu g^{\text{EH}}_{\mu\bnu}d\overline{\omega^\nu}-c.c) + (\omega^3 d\overline{\omega^3}+\omega^4
  d\overline{\omega^4}-c.c)\right]\,dx^+
 \label{eq:ineh}
\end{multline}
Notice that $F_5=dx^+\wedge \varphi_4$ keeps its form, just by replacing all the dependence in $z^N$ in $\varphi_4$ by $\omega^N$.
Let us introduce polar coordinates to describe both $\bC^2$ complex planes. In particular, consider
\begin{equation}
  \begin{aligned}[m]
    w^1 &=  \rho \cos (\theta/2)e^{\frac{i}{2}(\psi+\phi)}\;\;;\;\; w^2=  \rho \sin (\theta/2)e^{\frac{i}{2}(\psi-\phi)}~,\\
    w^3 &=  \tilde{\rho}\cos (\tilde{\theta}/2)e^{\frac{i}{2}(\tilde{\psi}+\tilde{\phi})}\;\;;\;\;  w^4=  \tilde{\rho} \sin (\tilde{\theta}/2)
    e^{\frac{i}{2}(\tilde{\psi}-\tilde{\phi})}~,
  \end{aligned}
 \label{eq:polcoors}
\end{equation}
where, as before, $r^4 = \rho^4+a^4$. In this way, we can rewrite the metric \eqref{eq:ineh} using the standard $\fsu(2)$ left invariant
one-forms $\{\sigma_i,\,\tilde{\sigma}_i\}$ as follows
\begin{equation}
  g =-2dx^+dx^{'-} +\|\xi\|^2\,(dx^+)^2  + g_{\text{EH}} + g_{\bC^2} + 2\nu \left(\tilde{\rho}^2\tilde{\sigma}_z
  + \frac{\rho^4}{r^2}\sigma_z\right)\,dx^+~,
 \label{eq:feh}
\end{equation}
where $g_{\text{EH}}=\frac{\rho^2}{r^2}(d\rho^2+\rho^2\sigma_z^2)+r^2(\sigma_x^2+\sigma_y^2)$ , 
$g_{\bC^2} = d\tilde{\rho}^2+\tilde{\rho}^2(\tilde{\sigma}_z^2+\tilde{\sigma}_x^2+\tilde{\sigma}_y^2)$ and
the norm of the Killing vector is given by
$\|\xi\|^2 =  - \{2a+(32\mu^2-\nu^2)\left(\tilde{\rho}^2+\frac{\rho^4}{r^2}\right)\}$.

The identification, $x^+\sim x^++2\pi R$ will create CTCs, as $\xi$ involves a compact rotation isometry.
This is more manifest in the T-dual description. Indeed, consider the T-dual configuration along the $\partial_{x^+}$ direction,
for the particular choice $32\mu^2=\nu^2$ and $a=-\frac{1}{2}$. This has a non-trivial NS-NS two-form potential,
$B_2 = \left(-dx^{'-}+\nu (\tilde{\rho}^2\tilde{\sigma}_z+ \frac{\rho^4}{r^2}\sigma_z)\right)\wedge dx^+$ and a metric
\begin{equation}
  g =-\left[dx^{'-} -\nu \left(\tilde{\rho}^2\tilde{\sigma}_z +
  \frac{\rho^4}{r^2}\sigma_z\right)\right]^2 + (dx^+)^2 + g_{\text{EH}} + g_{\bC^2}~.
 \label{eq:dualmet}
\end{equation}
We are ignoring here the details concerning a non-trivial RR four-form field strength.

By inspection of \eqref{eq:dualmet}, we see immediately that we have CTCs. Indeed, $g_{\psi\psi}= \frac{\rho^4}{4r^2}(1-\frac{\nu^2\rho^4}{r^2})$ and
$g_{\tilde{\psi}\tilde{\psi}}=\frac{\tilde{\rho}^2}{4}(1-\nu^2\,\tilde{\rho}^2)$, both becoming negative as $\rho,\tilde{\rho}$ increase.
Notice that the metric \eqref{eq:dualmet} can be regarded as a G\"{o}del-type universe over an Eguchi--Hanson space, and by construction
it is supersymmetric.

\subsubsection{D-branes in pp-wave backgrounds with transverse flat space}

As a final example of a supersymmetric background having a null isometry we will consider that of a joint (2,2) supersymmetric
pp-wave and a D-brane in type IIB theory. Such a background is not of the pp-wave type, as it has a non-trivial warp factor describing
the backreaction of the D-brane. Yet, it is easy to handle as its supersymmetries are known. When there is no D-brane, we know these pp-waves
have at least (2,2) supersymmetry. In the case on which we shall concentrate, the addition of the D-brane projects out half of these,
leaving us with a (2,0) or (0,2) supersymmetric background. We actually prove that the pp-wave superimposed
with a D1-brane in the $x^+,x^-$ directions preserves (2,0) supersymmetry. Using U-dualities one can obtain other supergravity
backgrounds in type IIB, type IIA and M-theory, where the pp-wave is superimposed with a brane in a way that preserves (2,0) supersymmetry.

Supergravity solutions for D-branes in plane wave backgrounds with flat transverse space have been previously obtained in
\cite{planeD}, and for D-branes in pp-wave backgrounds with flat transverse space such solutions were obtained
in \cite{DppAlday,DppSanjay}. Note that our solutions in appendix D differ from those obtained in \cite{DppAlday}
as they claim to get D1+pp solutions which are not supersymmetric, while we find (2,0) and (1,0)
supersymmetric solutions for the D1+pp system. Our results also differ from those of \cite{DppSanjay} as they try to find
supersymmetric solutions for a general flux, while we pick a flux related to the
metric, so that the entire background preserves some supersymmetry.

Even though our results are more general, let us work with the (2,0) supersymmetric pp wave+D1 background in flat transverse space.
The complete background is given by
\begin{equation}
  \begin{aligned}[m]
    g &= f^{-1/2}[-2dx^-dx^+ -32(|dW(z)|^2 +|V(z)|^2 )(dx^+)^2] +f^{1/2}dz^{\alpha}d{\bar z}^{\bar \alpha}~,\\
    \varphi_{\alpha\beta} &= \partial_{\alpha}\partial_{\beta}W ~,~~~~~~~~~~~~ \varphi_{\bar \alpha\bar \beta} = \partial_{\bar \alpha}
    \partial_{\bar \beta} \bar W ~,~~~~~~~~~~ \varphi_{\bar \alpha \beta} =\partial_{\bar \alpha} \partial_\beta U~, \\
    F_5 &= dx^+\wedge \varphi_4~, \\
    F_3 &= dx^+\wedge dx^-\wedge df^{-1}~,\\
    e^\phi &= g_s\,f^{1/2}~,
  \end{aligned}
 \label{eq:curi}
\end{equation}
where $f(z^\alpha,z^{\bar{\alpha}})$ is a real harmonic function, $W(z)$ a holomorphic function and $V^{\mu}$
a holomorphic Killing vector, such that $\partial_{\alpha}V^{\alpha}=\partial_{\alpha}[V^{\beta}\partial_\beta W ]
=0 $, $V_{\mu}=i\partial_{\mu}U$.

In the absence of the D-strings, we showed in Appendix C that any Killing vector of the form
\[
  \xi=a\partial_++b\partial_-+\tilde{V}~,
\]
preserves all the (2,2) supersymmetries if $\tilde{V}$ is a holomorphic Killing vector satisfying $\nabla_\mu\tilde{V}^\mu=0$ and 
\[
  {\cal L}_{\tilde{V}}V^\mu={\cal L}_{\tilde{V}}\nabla_{\mu}W=0~.
\]
In the presence of the D-strings, it is easy to see that the previous $\xi$ will preserve all the supersymmetries of the
joint ppwave+D1 set-up if it satisfies the further requirement
\[
  {\cal L}_{\tilde{V}}f=\tilde{V}^\mu\nabla_{\mu}f+\tilde{V}^{\bmu}\nabla_{\bmu}f=0~,
\]
which again ensures it is an isometry of the background \eqref{eq:curi}.

Under these circumstances, both $V$ and $\tilde{V}$ generate rotations in the transverse space,
\[
  V=i\,R^\mu_{\;\nu}z^\nu\,\partial_\mu\, +\,c.c. \quad , \quad \tilde{V}=i\,\tilde{R}^\mu_{\;\nu}z^\nu\,\partial_{\mu}\,+\,c.c.~.
\]
Therefore, if $\tilde{V}$ satisfies the conditions
\[
  [R,\tilde{R}]=0\;\; , \;\; \tilde{R}^{\mu}_{\;\nu}z^{\nu}\partial_{\mu}W=const~,
\]
it will be an isometry of \eqref{eq:curi} and will preserve all its supersymmetries.

Instead of dealing with the most general possibility, we shall focus on a particular example
in which we set $W=0$, $\sqrt{32}\,V^\mu=\tilde{V}^\mu = i\beta\,(z^1\partial_1-z^2\partial_2+z^3\partial_3-z^4\partial_4)$, $a=-2b=1$ and we take all
the D-strings to be localized at the origin, so that the real harmonic function is $f=1+Q/r^6$.
Therefore, we are dealing with the background
\begin{equation}
  \begin{aligned}[m]
    g &= f^{-1/2}[-2dx^-dx^+ -\beta^2r^2(dx^+)^2] +f^{1/2}dz^{\alpha}d{\bar z}^{\bar \alpha}~, \\
    \varphi_{\mu\nu} &= \varphi_{\bmu\bnu} = 0 \;\;;\;\; \varphi_{\bmu \nu} = \partial_{\bmu}
    \partial_{\nu}(\frac{\beta}{\sqrt{32}}(|z^1|^2-|z^2|^2+|z^3|^2-|z^4|^2))~, \\
    F_5 &= dx^+\wedge \varphi_4 \;\;\;,\;\;\; F_3 = dx^+\wedge dx^-\wedge df^{-1}~,\\
    e^\phi &= g_s\,f^{1/2} \quad , \quad  f = 1+\frac{Q}{r^6}~,
  \end{aligned}
 \label{eq:curii}
\end{equation}
and the generator of the discrete quotient that we shall analyse is given by
$\xi = \partial_+-\frac{1}{2}\partial_-+\beta(iz^1\partial_1-iz^2\partial_2+iz^3\partial_3-iz^4\partial_4+c.c.)$.

By changing coordinates to an adapted coordinate system defined by
\begin{equation*}
  x^{-'}=x^-+\frac{1}{2}x^+ \;,\; y^1=e^{-i\beta x^+}z^1 \;,\; y^2=e^{+i\beta x^+}z^2\;,\; y^3=e^{-i\beta x^+}z^3 \;,\; y^4=e^{i\beta x^+}z^4~,
\end{equation*}
in which $\xi=\partial_+$, the metric in \eqref{eq:curii} is
\begin{multline}
  g = f^{-1/2}(1+\beta^2Q^2/r^4)[dx^+ -(1+\beta^2Q^2/r^4)^{-1}(dx^{-'}-\beta f\sum_j r_j^2d\theta_j)]^2 + \\
  - \frac{f^{-1/2}}{1+\beta^2Q/r^4}[dx^{-'}-\beta f\sum_j r_j^2d\theta_j]^2 + f^{1/2}\sum_j[dr_j^2+r_j^2d\theta_j^2]~,
 \label{eq:DppKK}
\end{multline}
where we introduced polar coordinates in each of the $\bR^2$ planes, i.e. $y^j=r_j e^{i\theta_j}$ $j=1,2,3,4$.

In this form, it is obvious that orbifolding by $\xi$ would generate a spacetime with CTCs. For instance, the T-dual metric along $x^+$ has future directed CTCs given by the orbits of the vector field $-\partial_{\theta^1}$ for
$\beta\,r_1 > 1+\beta^2Q^2\frac{r^2-r_1^2z}{r^6}$.

\section*{Acknowledgements}

JS would like to thank O. Aharony, N. Drukker, B. Fiol, B. Kol, J.M. Figueroa-O'Farrill and M. Rozali, whereas LM
would like to thank E. Gimon, V. Hubeny, J. Maldacena, M. Rangamani, S. Ross and A. Strominger, for useful
discussions. JS would like to thank the Institute for Advanced Studies in Princeton, the Perimeter Institute and
the Aspen Center for Physics and LM would like to thank the Institute for Advanced Studies in Princeton, Princeton
University, The Weizmann Institute of Science and the organizers of Strings 2003 in Kyoto, for hospitality during
the different stages involved in this project. JS was supported by a Marie Curie Fellowship of the European
Community programme ``Improving the Human Research Potential and the Socio-Economic Knowledge Base'' under the
contract number HPMF-CT-2000-00480, during the initial stages of this project, by the Phil Zacharia fellowship
from January to August in 2003 and since then by the United States Department of Energy under the grant number
DE-FG02-95ER40893. JS travelling budget was also supported in part by a grant from the United States--Israel
Binational Science Foundation (BSF), the European RTN network HPRN-CT-2000-00122 and by Minerva, during the
initial stages of this project. JS would also like to thank the IRF Centers of Excellence program.
LM would like to thank Stichting FOM for support.

\newpage

\appendix

\section{Future directed timelike vectors}
\label{sec:appA}

In this appendix we will show that if a timelike vector $y$ is future directed with respect to a globally defined vector field $\tau$, it will
also be future directed with respect to any other future directed timelike vector $\tilde{\tau}$. In other words, the future directness
of a given timelike vector field $y$ is independent of the choice of the representative of the class of mutually future directed globally
defined timelike vector fields $\{\tau\}$.

Mathematically, what we want to show is that if the following conditions are satisfied
\begin{eqnarray}
  &\|y\|^2& <0\,\,,\,\,\|\tau\|^2<0\,\,,\,\, g(y\,,\tau) <0 \\
  &\|\tilde{\tau}\|^2 & <0\,\,,\,\, g(\tilde{\tau}\,,\tau) <0~,
\end{eqnarray}
this implies that
\begin{equation}
  g(y,\,\tilde{\tau})<0 .
\end{equation}

We will first prove this statement in Minkowski spacetime, and afterwards that it also applies to any spacetime with the same signature.
In Minkowski spacetime, from the condition that both $y$ and $\tau$ are timelike, we derive the condition
\begin{equation*}
  (y_0)^2\,(\tau_0)^2 > |\vec{y}|^2\,|\vec{\tau}|^2 > \left(\vec{y}\cdot \vec{\tau}\right)^2~,
\end{equation*}
where we split $y=(y_0,\,\vec{y})$, $\tau=(\tau_0,\,\vec{\tau})$ and $\vec{x}\cdot\vec{z}$ stands for the euclidean scalar product
of two vectors $\vec{x}$ and $\vec{z}$.

Due to the fact that $y$ is future directed, the condition $y_0\,\tau_0 > \vec{y}\cdot\vec{\tau}$ is satisfied.
This allows us to infer that
\[
  y_0\,\tau_0 > 0~.
\]
An analogous conclusion is obtained for the product $\tau_0\,\tilde{\tau}_0 > 0$.

If $\tilde{\tau}_0 > |\vec{\tilde{\tau}}|$, we conclude that both $\tau_0$ and $y_0$ are positive, and consequently, $y_0 > |\vec{y}|$.
If $\tilde{\tau}_0 < -|\vec{\tilde{\tau}}|$, we conclude that both $\tau_0$ and $y_0$ are negative, and consequently, $y_0 < -|\vec{y}|$.
In either case,
\begin{equation*}
  y_0\,\tilde{\tau}_0 > |\vec{y}|\,|\vec{\tilde{\tau}}| > \vec{y}\cdot \vec{\tilde{\tau}}~.
\end{equation*}
Thus, indeed $y$ is future directed with respect to the globally defined timelike vector field $\tilde{\tau}$.

To deal with the curved spacetime, we just notice that we can always introduce a local orthonormal frame in which our spacetime
is Minkowski, and where our previous proof applies. Since we assumed that our spacetimes are time-orientable, this means that the distinction
between future and past directed vectors can be done at any point, and therefore the local analysis of our metric is enough for our purposes.

\section{CTCs in a dimensionally reduced spacetime with spacelike isometry}
\label{sec:appB}

The purpose of this appendix is to show that given a ten dimensional spacetime with a compact and periodic spacelike isometry $\partial_z$
having a future directed closed timelike curve, this necessarily implies the existence of a future directed closed timelike curve
in the nine dimensionally reduced metric $\tilde{g}$.

Thus, we have a ten dimensional metric of the form \eqref{eq:space} and a globally defined timelike vector field $\tau$,
with respect to which a closed timelike curve $\{z(\lambda),\,x^i(\lambda)\}$ is future directed. And we want to show that there exists
a closed timelike curve $\tilde{x}^i(\lambda)$ which is future directed with respect to a globally defined nine dimensional timelike
vector field $\tilde{\tau}$.

We shall prove that the simple choice $\tilde{\tau}=\tau^i\partial_i$ and $\tilde{x}^i(\lambda)=x^i(\lambda)$ already fulfills
all the above requirements, where we decomposed $\tau=\tau^z\partial_z + \tau^i\partial_i$. All these requirements are trivial except for the
future directed character of $\tilde{x}^i(\lambda)$. Let us assume that given these choices
the nine dimensional curve is past directed at some point $\lambda_0$. We shall show that we get a contradiction.

Let us introduce the parameters $\alpha\equiv \Delta(\dot{z}+A\cdot \dot{x})$ and $\beta\equiv \Delta(\tau^z+A\cdot \tau)$, where dot stands for $\frac{d}{d\lambda}$ and where all expressions are evaluated at $\lambda_0$.
In terms of these, our ten dimensional assumptions can be written as
\[
  0 < \alpha^2 < -\|\dot{x}\|^2\,\, ,\,\, 0 < \beta^2 <- \|\tau\|^2\,\,,\,\, \alpha\beta < -\tilde{g}_9(\tau,\,\dot{x}) <0~,
\]
where $\|y\|^2=\tilde{g}_9(y,\,y)$ $\forall\,y$. These inequalities imply that
\[
\tilde{g}_9(\dot{x},\,\tau)^2 < \alpha^2 \beta^2 < (-\|\dot{x}\|^2)(-\|\tau\|^2)~.
\]
However, the last inequality can never be true for two timelike vectors and a metric of lorentzian signature, i.e. for any two timelike
vectors with components $a^i\,,b^i$ the following is always true:
\begin{equation}
  g_{ij}g_{kl}a^ib^k(a^lb^j-a^jb^l)\geq 0 ~.
 \label{eq:9bound}
\end{equation}
One way to show this is to work in a local orthonormal frame. This amounts to computing the left hand side of the above expression by
replacing $g_{ij}\to\eta_{ij}$ and expressing both vectors in the corresponding orthonormal frame basis. The result,
\begin{multline}
  g_{ij}g_{kl}a^ib^k(a^lb^j-a^jb^l) = (a\cdot b)^2 - \|a\|^2\,\|b\|^2 = \\
  (|\vec{a}|^2|\vec{b}|^2)\cos^2\theta_{ab}-2a_0b_0|\vec{a}||\vec{b}|\cos\theta_{ab}+(a_0^2|\vec{b}|^2+b_0^2|\vec{a}|^2-|\vec{a}|^2|\vec{b}|^2)~,
\end{multline}
is a parabola, when viewed as a function of the cosine of the local angle between both vectors, $\cos\theta_{ab}$. As a parabola, it
reaches its minimum at $\cos\theta_{ab}=\frac{a_0b_0}{|\vec{a}||\vec{b}|}$. Since $|\cos\theta_{ab}|\leq 1$, the minimum is obtained whenever
$\cos\theta_{ab}=\pm 1$. For these extremal values the entire expression becomes a perfect square
\begin{equation}
   (a\cdot b)^2 - \|a\|^2\,\|b\|^2 = (a_0|\vec{b}|\pm b_0|\vec{a}|)^2\geq 0~.
\end{equation}
Thus, we conclude that \eqref{eq:9bound} is always non-negative, and therefore the 9-dimensional curve is everywhere future directed.

\section{Supersymmetry analysis}
\label{sec:appC}

In this appendix we will look at the (2,2) supersymmetric pp-wave backgrounds described in \cite{juanpp} and
analyse which Killing vectors preserve all or some of their supersymmetries when considered as generators
of discrete quotients.

The (2,2) supersymmetric pp-wave backgrounds can be parameterized in the following way:
\begin{eqnarray}
  ds^2 &=& -2dx^+dx^- -32(|dW|^2+|V|^2)(dx^+)^2+2g_{\mu\bar{\nu}}dz^{\mu} d{\bz}^{\bnu} \nonumber \\
  F_5 &=& dx^+\wedge \varphi_4 \nonumber\\\varphi_{\mu\nu}&=&\nabla_\mu\nabla_\nu W ~~,~~ \varphi_{\bmu\bnu}=\nabla_{\bmu}\nabla_{\bnu}\bar{W} ~~,~~
  \varphi_{\bmu\nu}=\nabla_{\bmu}\nabla_{\nu}U
 \label{eq:liatmet}
\end{eqnarray}
where $W$ is a holomorphic function of the transverse coordinates and $V^\mu$ a holomorphic Killing vector derived
from a real Killing potential $U$, such that $\nabla_\mu V^\mu=0$ and $\partial_\nu[V^\mu\nabla_\mu W]=0$, and the four-form $\varphi_4$ is given
in terms of $\varphi_2$ by
\begin{equation}
  \varphi_4=\frac{1}{4!}[\varphi_{\mu\nu}g^{\nu\bar{\nu}}\epsilon_{\overline{\rho\sigma\lambda\nu}}dz^{\mu}\wedge d\bar{z}^{\bar{\rho}}\wedge
  d\bar{z}^{\bar{\sigma}}\wedge d\bar{z}^{\bar{\lambda}} +
  \frac{1}{2}\varphi_{\mu\bar{\nu}}g_{\rho\bar{\rho}}dz^{\mu}\wedge d\bar{z}^{\bar{\nu}}\wedge dz^{\rho}\wedge d\bar{z}^{\bar{\rho}}+c.c.]
 \label{eq:fourform}
\end{equation}

By computing the anticommutator of two supersymmetry generators we can find some bosonic isometries
of the background, confirming the Killing character of the holomorphic vector field $V^\mu$.
The spacetime Killing spinors are of the form (using the notations of \cite{juanpp}):
\begin{equation}
  \varepsilon = [\alpha+\frac{1}{4}\zeta \frac{1}{4!}\varepsilon_{\overline{\mu\nu\rho\sigma}}
  b^{\dag\bmu}b^{\dag\bnu}b^{\dag\bar{\rho}} b^{\dag\bar{\sigma}}]|0> +
  \Gamma_-[\beta_{\bmu}b^{\dag\bmu}+\delta_{\mu}\widetilde{b}^{\mu}]|0>~,
 \label{eq:epsi}
\end{equation}
where for $N=2$ solutions we have
\begin{equation}
  \beta_{\bar{\mu}}=2i[i\alpha V_{\bar{\mu}}+\zeta\overline{\nabla_{\mu}W}] ~~;~~ \delta_{\mu}=2i[-i\zeta
  V_{\mu}+\alpha\nabla_{\mu}W]~.
 \label{eq:mtwo}
\end{equation}
The Gamma matrices are all real, $\Gamma^0$ is skew-hermitian $(\Gamma^0)^\dag=-\Gamma^0$ and
$\Gamma^1,...\Gamma^8$ are all hermitian.
The charge conjugation matrix is $C=\Gamma^0$ which obeys $\Gamma_{\pm}C=C\Gamma_{\mp}$ and anticommutes
with all other $\Gamma^1,...,\Gamma^9$. Denoting $<0|\equiv |0>^\dag$, the hermitian conjugate of \eqref{eq:epsi} is
\begin{eqnarray}
  \bar{\varepsilon}=\varepsilon^\dag C &= <0|[\alpha^*+\zeta^*\frac{1}{4} \frac{1}{4!}\varepsilon_{\mu\nu\rho\sigma}
  b^{\mu}b^{\nu}b^{\rho} b^{\sigma}]C-<0|[\beta^*_\mu b^\mu+\delta^*_{\bmu}\frac{1}{4}
  \varepsilon_{\lambda\nu\rho\sigma}\frac{1}{4!}b^\lambda b^\nu b^\rho b^\sigma b^{\bmu}]\Gamma_+C~.
 \label{eq:epsibar}
\end{eqnarray}
We therefore find that
\begin{eqnarray}
  {\cal V} &=&\bar{\varepsilon}_1\Gamma^{\rho}\varepsilon_2\partial_{\rho} +
  \bar{\varepsilon}_1\overline{\Gamma^{\rho}}\varepsilon_2\overline{\partial_{\rho}} +
  \bar{\varepsilon}_1\Gamma^+\varepsilon_2\partial_++\bar{\varepsilon}_1\Gamma^-\varepsilon_2\partial_-=
  \nonumber\\ &=& <0|C\Gamma_-|0>\{2[(\alpha_1^*\beta_2^\mu +\zeta_2\delta_1^{*\mu})\partial_\mu
  +(\alpha_2\beta_1^{*\bmu}+\delta_2^{\bmu} \zeta_1^*)\partial_{\bmu}] + \nonumber \\
  &-&(\alpha_1^*\alpha_2+\zeta_1^*\zeta_2)\partial_+ -[H(\alpha_1^*\alpha_2+\zeta_1^*\zeta_2)
  +4(\beta_1^*\beta_2+\delta_1^*\delta_2)]\partial_-\} ~.
 \label{eq:calc}
\end{eqnarray}
By plugging in the expressions for $\beta_{\mu} , \delta_{\mu}$ from \eqref{eq:mtwo}, all the
dependence on $\nabla W$ in the transverse piece cancels and one is left with
\begin{eqnarray}
  {\cal V} &=& -4(\alpha_1^*\alpha_2+\zeta_1^*\zeta_2)(V^{\mu}\partial_{\mu}
  +V^{\bar{\mu}}\overline{\partial_{\mu}})<0|C\Gamma_-|0> \nonumber\\
  &+&[-(\alpha_1^*\alpha_2+\zeta_1^*\zeta_2)\partial_u+32i(\zeta_1^*\alpha_2V^\mu\nabla_\mu
  W-\alpha_1^*\zeta_2V^{\bmu}\overline{\nabla_\mu W})\partial_v]<0|C\Gamma_-|0> ~.
 \label{eq:calcb}
\end{eqnarray}
To derive the above expression, we used $\partial_u=\partial_++(H/2) \partial_-$ and
$\partial_v=\partial_-$. Note that the last term proportional to $\partial_v$
is just a constant.

Thus, the anticommutator of any two supersymmetries is proportional to the Killing vectors of the space:
$V,\partial_u,\partial_v$ (the generators associated with the later two are
usually denoted as $p_+\,,p_-$). We would like to stress that it is still possible that there might be other
bosonic symmetries which commute with the supercharges, but that do not appear in the anticommutator of two of them.

In the following, we shall analyse which subset of Killing vectors $(\xi)$ leave all or some of the Killing
spinors \eqref{eq:epsi} invariant. By the philosophy explained at the beginning of section \ref{sec:juanliat}, the
discrete quotients generated by identifying points under a discrete step transformation generated by $\xi$ will
preserve the same amount of supersymmetry as the number of preserved Killing spinors (at least locally). One is
therefore instructed to solve the equation $\cL_\xi\varepsilon=0$ \cite{lie}, where
\begin{eqnarray}
  \cL_\xi \varepsilon &=& \xi^M\nabla_M\varepsilon+\frac{1}{4}\nabla_{M}\xi_{N}\Gamma^{MN}\varepsilon = \nonumber \\
  &=&\frac{1}{4}\xi^+ \sp H\Gamma_-\varepsilon +\xi^{\mu}\nabla_{\mu}\varepsilon+\xi^{\bmu}\nabla_{\bmu}\varepsilon+ \nonumber \\
  &+&\frac{1}{4}[(\partial_+\xi_{\alpha}-\partial_{\alpha}\xi_+)\Gamma^{+\alpha}+
  (\partial_-\xi_{\alpha}-\partial_{\alpha}\xi_-)\Gamma^{-\alpha}+(\partial_+\xi_- -\partial_-\xi_+)\Gamma^{+-}
  +\partial_{\alpha}\xi_{\beta}\Gamma^{\alpha\beta}]\varepsilon ~.
 \label{eq:Leps}
\end{eqnarray}
In the above equation, $M,N=+,-,\mu,\bmu$ and $\alpha,\beta$ stand for both holomorphic and anti-holomorphic indices.
Although we could proceed with a general analysis, we shall focus on vectors fields $\xi$ which are independent of $x^\pm$ :
$\partial_\pm \xi^M=0$ and which have constant $\xi^\pm$ components: $\partial_\alpha\xi^\pm=0$.
Under these assumptions, equations \eqref{eq:Leps} simplify: the first term in the
second line cancels the first term in the third line of \eqref{eq:Leps}, the second and third terms of the third
line vanish, so that one is left with the equation
$$
  \cL_\xi\varepsilon =\xi^{\alpha}\nabla_{\alpha}\varepsilon+\frac{1}{4}\partial_{\alpha}\xi_{\beta}\Gamma^{\alpha\beta}\varepsilon=0~.
$$
Using the exact expression for the Killing spinor $\varepsilon$ given in \eqref{eq:epsi}-\eqref{eq:mtwo} one can derive the
following set of identities :
\begin{eqnarray}
  \nabla_{\lambda}\varepsilon &=& 2i\Gamma_-[i\alpha \nabla_{\lambda}V_{\bmu}b^{\dag\bmu}+\alpha\nabla_{\lambda}\nabla_{\mu}W\widetilde{b}^{\mu}]|0>
  ~,\nonumber \\
  \nabla_{\bar{\lambda}}\varepsilon &=&   2i\Gamma_-[\zeta\nabla_{\bar{\lambda}}\overline{\nabla_{\mu}W}b^{\dag\bmu}-
  i\zeta \nabla_{\bar{\lambda}}V_{\mu}\widetilde{b}^{\mu}]|0>~,
\end{eqnarray}
and
\begin{eqnarray}
  \Gamma^{\lambda\bar{\lambda}}\varepsilon &=& g^{\lambda\bar{\lambda}} [\alpha-\frac{1}{4}\zeta
  \frac{1}{4!}\varepsilon_{\overline{\mu\nu\rho\sigma}} b^{\dag\bmu}b^{\dag\bar{\nu}}b^{\dag\bar{\rho}}b^{\dag\bar{\sigma}}]|0>+ \nonumber \\
  &+&  2i\Gamma_-[(\zeta\overline{\nabla_{\mu}W}+i\alpha V_{\bmu})(g^{\lambda\bar{\lambda}}b^{\dag\bmu}-2g^{\bmu\lambda}b^{\dag\bar{\lambda}})
  +(\alpha\nabla_{\mu}W-i\zeta V_{\mu})(-g^{\lambda\bar{\lambda}}\widetilde{b}^{\mu}+2g^{\mu\bar{\lambda}}\widetilde{b}^{\lambda})]|0>~,
  \nonumber \\
  \Gamma^{\bar{\lambda}\bar{\omega}}\varepsilon &=& \alpha b^{\dag\bar{\lambda}}b^{\dag\bar{\omega}}|0> + 2i\Gamma_-(\zeta\overline{\nabla_{\mu}W}+i\alpha
  V_{\bmu})(-2\varepsilon^{\overline{\lambda\omega\mu\nu}})g_{\bnu\nu}\widetilde{b}^{\nu}]|0>~, \nonumber \\
  \Gamma^{\lambda\omega}\varepsilon &=& -\frac{1}{2}\zeta \varepsilon^{\lambda\omega\rho\sigma}g_{\rho\bar{\rho}}g_{\sigma\bar{\sigma}} b^{\dag\bar{\rho}}
  b^{\dag\bar{\sigma}}|0> +2i\Gamma_-(\alpha\nabla_{\mu}W-i\zeta V_{\mu})(-2\varepsilon^{\mu\lambda\omega\sigma})
  g_{\sigma\bar{\sigma}}b^{\dag\bar{\sigma}}|0>~.
\end{eqnarray}
We shall look for solutions of $\cL_\xi\varepsilon=0$ by plugging in the above identities.
First, consider the terms not proportional to $\Gamma_-$. From the coefficients of $|0>$ and $(b^\dag)^4|0>$, we learn that
$g^{\lambda\bar{\lambda}}\nabla_{[\lambda}\xi_{\bar{\lambda}]}=0$ has to be satisfied, whereas from the vanishing of the $(b^\dag)^2|0>$
coefficient, one gets
$\zeta\nabla_{\mu}\xi_{\nu}+\alpha\frac{1}{2!}\nabla_{\bar{\rho}}\xi_{\bar{\sigma}}\varepsilon_{\mu\nu}^{~~\bar{\rho}\bar{\sigma}}=0$.

Second, let us analyse the $\Gamma_-$ proportional pieces. The coefficient of the $b^\dag|0>$ terms is given by
\begin{multline*}
  0 = i\alpha \,\xi^{\lambda}\nabla_{\lambda}V_{\bmu}+\zeta \xi^{\bar{\lambda}}\overline{\nabla_{\lambda}\nabla_{\mu}W} \\
  + \frac{1}{4}\{ 2\nabla_{[\lambda}\xi_{\bar{\lambda}]}(\zeta\overline{\nabla_{\mu}W}+i\alpha V_{\bmu})g^{\lambda\bar{\lambda}}
  -4\nabla_{[\lambda}\xi_{\bar{\mu}]}(\zeta\overline{\nabla_{\sigma}W}+i\alpha
  V_{\bar{\sigma}})g^{\bar{\sigma}\lambda} -2\nabla_{[\lambda}\xi_{\omega]}(\alpha\nabla_{\sigma}W-i\zeta
  V_{\sigma})\varepsilon^{\sigma\lambda\omega\mu} g_{\mu\bar{\mu}}\}~.
\end{multline*}
Note that in the second line above, the first and second terms get a $2$ factor, as we did not write down the
$\nabla_{[\bar{\lambda}}\xi_{\lambda]}$ term. The first term on the second line vanishes as we concluded from the previous
discussion. The last term vanishes when $|\zeta|\neq |\alpha|$, so that $\xi$ is holomorphic (we discuss the
$|\alpha|=|\zeta|$ case below). Therefore, using holomorphicity of $V^\mu$, one gets the final constraint
\begin{equation*}
  i\alpha (\xi^{\lambda}\nabla_{\lambda}V_{\bmu} -\nabla_{[\lambda} \xi_{\bar{\mu}]}
  (V_{\bar{\sigma}})g^{\bar{\sigma}\lambda})+\zeta (\xi^{\bar{\lambda}}\overline{\nabla_{\lambda}\nabla_{\mu}W}
  -\nabla_{[\lambda}\xi_{\bar{\mu}]}g^{\bar{\sigma}\lambda}\overline{\nabla_{\sigma}W}) = i\alpha \cL_\xi V_{\bmu}+\zeta\cL_\xi\overline{\nabla_{\mu}W}=0~.
\end{equation*}
Similarly, the vanishing of the $\widetilde{b}^\mu$ terms gives, whenever $|\alpha|\neq |\zeta|$, the constraint
\begin{equation*}
  \alpha[\xi^\lambda\nabla_{\lambda}\nabla_{\mu} W + g^{\lambda\bar{\lambda}}\nabla_{[\mu}\xi_{\bar{\lambda}]}\nabla_{\lambda}W]
  -i\zeta[\xi^{\bar{\lambda}}\nabla_{\bar{\lambda}}V_{\mu}+\nabla_{[\mu}\xi_{\bar{\lambda}]}V^{\bar{\lambda}}] =\alpha\cL_\xi \nabla_{\mu}W-i\zeta\cL_\xi V_{\mu}=0~.
\end{equation*}
The above analysis allows us to conclude that if $|\zeta|\neq |\alpha|$, the vector fields satisfying equation \eqref{eq:susyquotient}
must satisfy
\begin{equation}
  \nabla_{\mu}\xi_{\nu}=0 \;,\; \nabla_{\mu}\xi^{\mu}=0 \;,\; \cL_\xi V^\mu = \cL_\xi \nabla_{\mu}W =0~.
 \label{eq:contwo}
\end{equation}
Equivalently, the vector field $\xi$ satisfies the same conditions as $V^\mu$ \footnote{Note that this is more general
than what we deduced previously from the algebra. $\xi$ has to be a holomorphic Killing vector, a symmetry of both $dW,V$ but does
not necessarily equal $V$.}. Notice that any transverse Killing vector field satisfying \eqref{eq:contwo}, also satisfies
$\cL_{\xi}g_{MN}=0$ and $\cL_{\xi}F_5=0$. Therefore, any vector field $\xi$ satisfying \eqref{eq:contwo} is a Killing vector
field, and as such, it can be used to construct a discrete quotient out of our original spacetime.

Finally, if $|\alpha|=|\zeta|$, one can consider the more  general solution with (1,1) supersymmetry. This is described by
the configuration
\begin{eqnarray}
  ds^2 &=& -2dx^+dx^- -32(|dU|^2)(dx^+)^2+2g_{\mu\bar{\nu}}dz^{\mu} d{\bz}^{\bnu} \nonumber\\ F_5 &=& dx^+\wedge \varphi_4 \nonumber\\
  \varphi_{\mu\nu}&=&\nabla_\mu\nabla_\nu U ~~,~~ \varphi_{\bmu\bnu}=\nabla_{\bmu}\nabla_{\bnu}U ~~,~~
  \varphi_{\bmu\nu}=\nabla_{\bmu}\nabla_{\nu}U~,
 \label{eq:(1,1)}
\end{eqnarray}
whereas their Killing spinors are given by
\begin{equation}
  \varepsilon = [1-\frac{1}{4}\frac{1}{4!}(\varepsilon_{\overline{\mu\nu\rho\sigma}}
  b^{\dag\bar{\mu}}b^{\dag\bar{\nu}}b^{\dag\bar{\rho}}b^{\dag\bar{\sigma}})]|0>
  -2i\Gamma_-[\nabla_{\bnu}Ub^{\dag\bnu}-\nabla_\nu U\tilde{b}^\nu]|0>~.
\end{equation}
In the above expressions, $U$ is a real harmonic function, and $\varphi_4$ is given in terms of $\varphi_2$ as in \eqref{eq:fourform}.
The same analysis as before, carried for this (1,1) configuration
gives rise to the following set of conditions on the vector field $\xi$
\begin{equation}
  \nabla_{\mu}\xi^{\mu}=0  \;,\; \nabla_{\bmu}\xi_{\bnu}+\frac{1}{2}\nabla_{\lambda}\xi_{\omega}\varepsilon^{\lambda\omega}_{~~\bmu\bnu}=0
  \;,\; \cL_\xi \nabla_{\mu}U =0~.
 \label{eq:conone}
\end{equation}
As before, these conditions along with the requirement that $\xi$ is a transverse Killing vector, guarantee that $\xi$ is a Killing vector field of the supergravity configuration \eqref{eq:(1,1)}.

We therefore establish that a vector field $\xi$ satisfying conditions \eqref{eq:contwo} (conditions \eqref{eq:conone}),
is necessarily an isometry of the bosonic background and preserves its (2,2) ((1,1)) supersymmetries.

\section{Supersymmetric backgrounds of pp-waves and D-branes}
\label{sec:appD}

\subsection{preliminaries: spin-connections and covariant derivatives}

We shall look for supergravity configurations whose metric ansatz is described by

\begin{equation}
  ds^2 = f^{-1/2}(y)[-2dx^+dx^-+S(x,y)(dx^+)^2+g_{\mu\nu}(x)dx^{\mu}dx^{\nu}] +
  f^{1/2}(y)g_{\alpha\beta}(y)dy^{\alpha}dy^{\beta}~.
\end{equation}
$g_{\mu\nu}(x)$ stands for an arbitrary euclidean (p-1)-dimensional metric parameterized by $\{x^\mu\}$, whereas
$g_{\alpha\beta}(y)$ describes a (9-p)-dimensional metric in coordinates $\{y^\alpha\}$. The scalar function
$f(y)$ was chosen to be independent of the $\{x\}$ coordinates, since those are going to be spacelike directions
along the worldvolume of the D-brane, besides the lightlike coordinates $\{x^\pm\}$.

We choose the \emph{vierbeins} ($ds^2 = -2\theta^u\theta^v +\theta^i\theta^i+ \theta^a \theta^a$) to be
\begin{eqnarray}
  \theta^u &=&  f^{-1/4}dx^+ \;\;\;,\;\;\; \theta^v = f^{-1/4}[dx^- -\frac{1}{2}Sdx^+] \nonumber\\
  \theta^i &=& f^{-1/4}e^i \;\;\;,\;\;\; \theta^a = f^{1/4}e^a
\end{eqnarray}
where $\{e^i\}$ stand for the vierbeins of the (p-1)-dimensional metric $g_{\mu\nu}(x)$, and $\{e^a\}$ are the
corresponding ones for the metric $g_{\alpha\beta}(y)$.

Given these vierbeins, the \emph{spin-connection} is determined to be
\begin{eqnarray}
  \Omega^u_{~a}&=&\Omega^a_{~v}=-\frac{1}{4}f^{-5/4}\partial_af \,\theta^u\, \nonumber\\
  \Omega^v_{~i}&=&\Omega^i_{~u}=-\frac{1}{2}f^{1/4}\partial_iS \,\theta^u\, \nonumber\\
  \Omega^v_{~a}&=&\Omega^a_{~u}=\frac{1}{4}f^{-5/4}\partial_af\,\theta^v -\frac{1}{2}f^{-1/4}\partial_aS\,\theta^u\, \nonumber\\
  \Omega^i_{~a}&=&-\Omega^a_{~i}=-\frac{1}{4}f^{-5/4}\partial_af\,\theta^i\, \nonumber\\
  \Omega^a_{~b}&=&\omega^a_{~b}+\frac{1}{4}f^{-5/4}\partial_bf\,\theta^a\, \nonumber\\
  \Omega^i_{~j}&=&\omega^i_{~j}\,
\end{eqnarray}
where $\partial_i = e_i^{\mu}\partial_{\mu}$ , $\partial_a = e_a^{\alpha}\partial_{\alpha}$, and $\omega$ is the spin-connection
of the transverse metric $g_{\alpha\beta}$. Out of the spin-connection, we can compute the action of the covariant
derivative on the gravitino field $\psi$. This action is given below
\begin{eqnarray}
  \nabla_+\psi &=& \partial_+\psi +\frac{1}{4}[-\frac{1}{4}f^{-3/2}\partial^af \Gamma_{ua}-\frac{1}{2}
  \partial^iS\Gamma_{vi}+(\frac{S}{8f^{3/2}}\partial^af-\frac{1}{2f^{1/2}}\partial^aS)\Gamma_{va}]\psi =\nonumber\\
  &=&\partial_+\psi +\frac{1}{4}[\frac{1}{4f^{3/2}}\sp f \Gamma_{u} -\frac{S}{8f^{3/2}}\sp f\Gamma_v +\frac{1}{2}
  (\Gamma^i\partial_iS +\frac{1}{f^{1/2}}\Gamma^a\partial_aS)\Gamma_{v}]\psi\, \nonumber\\
  \nabla_-\psi &=& \partial_-\psi +\frac{1}{4}\frac{1}{4f^{3/2}}\sp f\Gamma_v\psi\,\nonumber\\
  \nabla_{\mu}\psi &=& (\partial_{\mu}\psi +\frac{1}{4}\omega^{ij}_{\mu}\Gamma_{ij}\psi)\, \nonumber\\
  \nabla_{\alpha}\psi &=& (\partial_{\alpha}\psi +\frac{1}{4}\omega^{ab}_{\alpha}\Gamma_{ab}\psi) +\frac{1}{4}
  \frac{1}{4f}\partial^bfe^a_{\alpha}\Gamma_{ab}\psi~.
\end{eqnarray}

\subsection{pp+Dp-brane solutions}

As different pp+Dp brane solutions are related by U-dualities, we can work out an example for a specific $Dp$
background, and then generalize our results. It turns out the D-string analysis $(p=1)$ is particularly analogous
to the supersymmetry analysis in \cite{juanpp}. We shall focus on D-strings extending along the $+,-$ directions,
and to begin with, on a flat transverse metric $g_{\alpha\bar\beta}= \delta_{\alpha\bar\beta}$. The full ansatz
that we shall consider involves the $p=1$ metric introduced before, plus the necessary non-trivial dilaton and RR
three-form field strength taking care of the stabilization of the system and the RR charge carried by the
D-strings.
\begin{equation}
  \begin{aligned}[m]
    ds^2 &= f^{-1/2}(y)[-2dx^+dx^-+S(y)(dx^+)^2] +f^{1/2}(y)dy^{\alpha}dy^{\alpha}~, \\
    F_{3} &= dx^+\wedge dx^-\wedge df^{-1}~, \\
    F_5 &= dx^+\wedge \varphi_4 ~,\\
    e^{\phi}&=g_s\,\sqrt{f}~,
  \end{aligned}
 \label{eq:ansatzDi}
\end{equation}
where $\varphi_4$ is a closed and self-dual form in the 8-dimensional $y^{\alpha}$ space, and $f$ is a harmonic
function defined on the same space.

The dilatino and gravitino supersymmetry variations can be evaluated, using the expressions for the spin-connections
shown in the previous subsection,
\begin{eqnarray}
  \delta\chi &=& \frac{1}{2f^{5/4}}\partial_{\alpha}f\Gamma_{\underline{\alpha}}[1-\Gamma_{vu}\sigma^1]\varepsilon =0 \\
  \delta \Psi_{\alpha} &=& \partial_{\alpha}\varepsilon +\frac{1}{8f}\partial_{\beta}f(\Gamma_{\underline{\alpha\beta}}
  +\Gamma_{vu}\Gamma_{\underline{\beta}} \Gamma_{\underline{\alpha}}\sigma^1)\varepsilon
  +\frac{1}{16}\frac{1}{4!}\varphi_{\beta\gamma\delta\zeta}\Gamma^{\underline{\beta\gamma\delta\zeta}}
  \Gamma_vi\sigma_2\varepsilon=0 \\
  \delta\Psi_- &=&\partial_-\varepsilon+\frac{1}{8f^{3/2}}\partial_{\beta}H\Gamma_{\underline{\beta}}\Gamma_v(1-\sigma_1)\varepsilon \\
  \delta\Psi_+ &=& \partial_+\varepsilon +\frac{1}{8f^{3/2}}\partial_{\beta}H\Gamma_{\underline{\beta}}\Gamma_u(1+\sigma_1)\varepsilon
  -\frac{S\partial_{\beta}H}{16f^{3/2}}\Gamma_{\underline{\beta}}\Gamma_v(1-\sigma_1)\varepsilon +\frac{1}{4f^{1/2}}
  \partial_{\beta}S\Gamma_{\underline{\beta}}\Gamma_v\varepsilon \nonumber \\
  &+&\frac{1}{16\sqrt{f}}\frac{1}{4!}\varphi_{\beta\gamma\delta\zeta}\Gamma^{\underline{\beta\gamma\delta\zeta}}
  \Gamma_v\Gamma_ui\sigma_2\varepsilon =0~,
 \label{eq:susy}
\end{eqnarray}
where $\sigma^{1,2}$ are Pauli matrices acting on $\varepsilon$ as a doublet of sixteen complex component
spinors .

To solve these equations, we define a couple of chiral spinors $\varepsilon_\pm$ by
\begin{equation}
  f^{1/8}\varepsilon=-\frac{1}{2}\Gamma_{u}\Gamma_v\,f^{1/8}\varepsilon -\frac{1}{2}\Gamma_v\Gamma_u\,
  f^{1/8}\varepsilon \equiv \varepsilon_+ + \varepsilon_-~.
 \label{eq:epspm}
\end{equation}
The first three equations in \eqref{eq:susy} imply that $\varepsilon_+$ is a constant
spinor, and that
\[
  (1+\Gamma_{uv}\sigma_1) \varepsilon = (1+\sigma_1)\varepsilon_-=(1-\sigma_1)\varepsilon_+=0~.
\]
The last two equations in \eqref{eq:susy} become
\begin{equation}
  \begin{aligned}[m]
    \partial_\alpha \varepsilon_- &= -\frac{1}{16}\frac{1}{4!}\varphi_{\beta\gamma\delta\zeta}
    \Gamma^{\beta\gamma\delta\zeta}\Gamma_{\alpha}(\Gamma_vi\sigma_2\varepsilon_+) \\
    -4\,f^{1/2}\,i\sigma_2 \partial_+ &\varepsilon_- +\frac{1}{2}\frac{1}{4!} \varphi_{\beta\gamma\delta\zeta}
    \Gamma^{\beta\gamma\delta\zeta}\varepsilon_- = \partial_\beta S\Gamma_\beta (\Gamma_v i\sigma_2\varepsilon_+)~.
  \end{aligned}
\end{equation}

Notice  that if we assume $\partial_+\varepsilon_-=0$, then the harmonic function $f$ factors out completely from
the above  equations. Under these circumstances, they become exactly the supersymmetry equations one gets for a
background describing a pp-wave with no D-brane. Thus they are solved by a constant $\varepsilon_+$ which is
parameterized by two complex parameters in the case of (2,2) supersymmetry, or by one complex parameter in the
case of (1,1) supersymmetry, and by a chiral spinor  $\varepsilon_-$ which is completely determined in terms of
the other one $\varepsilon_+$.

By adding the requirement that $(1+\sigma_1)\varepsilon_-=0$ , we lose half
of the supersymmetries (as this condition will relate the real and imaginary parts of the complex
parameters). On the other hand, the condition $(1-\sigma_1)\epsilon_+=0$ would be automatically satisfied.
Therefore the D-string leaves us with only half of the supersymmetries preserved by the pp-wave.

We summarized the full supergravity solution for the (2,0) and (0,2), and for the (0,1) and (1,0) supersymmetric
configuration below, following the notations in \cite{juanpp}.

(*) (2,0) and (0,2) susy:
\begin{equation}
  \begin{aligned}[m]
    ds^2 &= f^{-1/2}[-2dx^-dx^+ -32(|dW(y)|^2 +|V(y)|^2 )(dx^+)^2] +f^{1/2}dy^{\alpha}d{\bar y}^{\bar \alpha}~, \\
   \varphi_{\alpha\beta} &= \partial_{\alpha}\partial_{\beta}W ~,~~~~~~~~~~~~
   \varphi_{\bar \alpha\bar \beta} = \partial_{\bar \alpha}\partial_{\bar \beta} \bar W ~,~~~~~~~~~~ \varphi_{\bar \alpha \beta}  =
   \partial_{\bar \alpha} \partial_\beta U~, \\
   F_5 &= dx^+\wedge \varphi_4 ~,\\
   F_3 &= dx^+\wedge dx^-\wedge df^{-1}~,\\
   e^\phi &= g_sf^{1/2}~,
  \end{aligned}
\end{equation}
where f is a real harmonic function, W a holomorphic function and $V^{\mu}$ a holomorphic Killing vector, such
that $\partial_{\alpha}V^{\alpha}=\partial_{\alpha}[V^{\beta}\partial_\beta W ] =0$ and $V_\alpha=i\partial_{\alpha}U$.

(*) (0,1) and (1,0) susy:
\begin{equation}
  \begin{aligned}[m]
    ds^2 &= f^{-1/2}[-2dx^-dx^+ -32( |d U|)^2 (dx^+)^2] +f^{1/2}dy^{\alpha}d\overline{y^{\alpha}}~,\\
    \varphi_{\alpha\beta} &= \partial_{\alpha}\partial_{\beta} U ~~;~~ \varphi_{\alpha\bar{\beta}} =\partial_{\alpha}\partial_{\bar{\beta}}U
    ~~;~~\varphi_{\bar{\alpha}\bar{\beta}} =\partial_{\bar{\alpha}\bar{\beta}}=\partial_{\bar{\alpha}}\partial_{\bar{\beta}}U~, \\
    F_5 &= dx^+\wedge \varphi_4~, \\
    F_3 &= dx^+\wedge dx^-\wedge df^{-1}~, \\
    e^\phi &= g_sf^{1/2}~,
  \end{aligned}
\end{equation}
$U$ , $f$ being real harmonic functions.

It is possible to extend the above solution to a curved transverse space $g_{\alpha\bar\beta}$.
The above configurations are generalised to :

(*) (2,0) and (0,2) susy:
\begin{equation}
  \begin{aligned}[m]
    ds^2 &= f^{-1/2}[-2dx^-dx^+ -32(|dW(y)|^2 +|V(y)|^2 )(dx^+)^2] +f^{1/2}g_{\alpha\bar{\beta}}dy^{\alpha}d\bar{y}^{\bar {\beta}}~,\\
    \varphi_{\alpha\beta} &= \nabla_{\alpha}\nabla_{\beta}W ~,~~~~~~~~~~~~
    \varphi_{\bar {\alpha}\bar {\beta}} = \nabla_{\bar {\alpha}}\nabla_{\bar{\beta}} \bar{W} ~,~~~~~~~~~~~~ \varphi_{\bar{\alpha} \beta} =
    \nabla_{\bar {\alpha}} \nabla_{\beta} U~, \\
    F_5 &= dx^+\wedge \varphi_4~, \\
    F_3 &= dx^+\wedge dx^-\wedge df^{-1}~,\\
    e^\phi &= g_sf^{1/2}~,
  \end{aligned}
\end{equation}
where $f$ is a real harmonic function, W a holomorphic function and $V^{\alpha}$ a holomorphic Killing
vector $V_{\alpha}=i\nabla_{\alpha} U$, such that $\nabla_{\alpha}V^{\alpha}=\nabla_{\alpha}
[V^{\beta}\nabla_\beta W ] =0$.

(*) (0,1) and (1,0) susy:
\begin{equation}
  \begin{aligned}[m]
    ds^2 &= f^{-1/2}[-2dx^-dx^+ -32( |d U|)^2 (dx^+)^2] +f^{1/2}g_{\alpha\bar{\beta}}dy^{\alpha}d\bar{y}^{\bar{\beta}}~, \\
    \varphi_{\alpha\beta} &= \nabla_{\alpha}\nabla_{\beta} U ~~;~~ \varphi_{\alpha\bar{\beta}} = \nabla_{\alpha}\nabla_{\bar{\beta}}U ~~;~~
    \varphi_{\bar{\alpha}\bar{\beta}} =\nabla_{\bar{\alpha}}\nabla_{\bar{\beta}}U~, \\
    F_5 &= dx^+\wedge \varphi_4~,\\
    F_3 &= dx^+\wedge dx^-\wedge df^{-1}~,\\
    e^\phi &= g_sf^{1/2}~,
  \end{aligned}
\end{equation}
where $U$ and $f$ are real harmonic functions. All covariant derivatives appearing above are taken with respect
to $g_{\alpha\bar{\beta}}$.

Once we have worked out the previous general solutions for D-strings in generalised pp-waves propagating
in curved backgrounds, it is possible to generate new solutions whenever there are enough symmetries
in the above configurations such that we can apply a T-duality transformation. Indeed, if we smear
all the functions and metric coefficients along some complex coordinates (and their complex conjugates), we shall generate
higher dimensional brane solutions. Combining this generating technique with S-duality and liftings to M-theory, we can
write different kinds of supersymmetric D-branes, F-strings, NS5-branes and M-branes solutions in these generalised pp-wave
backgrounds.

\end{document}